\PassOptionsToPackage{unicode}{hyperref}
\PassOptionsToPackage{hyphens}{url}
\documentclass[
]{article}
\usepackage{xcolor}
\usepackage[margin=1in]{geometry}
\usepackage{amsmath,amssymb}
\usepackage{iftex}
\ifPDFTeX
  \usepackage[T1]{fontenc}
  \usepackage[utf8]{inputenc}
  \usepackage{textcomp} 
\else 
  \usepackage{unicode-math} 
  \defaultfontfeatures{Scale=MatchLowercase}
  \defaultfontfeatures[\rmfamily]{Ligatures=TeX,Scale=1}
\fi
\usepackage{lmodern}
\ifPDFTeX\else
\fi
\IfFileExists{upquote.sty}{\usepackage{upquote}}{}
\IfFileExists{microtype.sty}{
  \usepackage[]{microtype}
  \UseMicrotypeSet[protrusion]{basicmath} 
}{}
\makeatletter
\@ifundefined{KOMAClassName}{
  \IfFileExists{parskip.sty}{%
    \usepackage{parskip}
  }{
    \setlength{\parindent}{0pt}
    \setlength{\parskip}{6pt plus 2pt minus 1pt}}
}{
  \KOMAoptions{parskip=half}}
\makeatother
\usepackage{longtable,booktabs,array}
\usepackage{calc} 
\usepackage{etoolbox}
\makeatletter
\patchcmd\longtable{\par}{\if@noskipsec\mbox{}\fi\par}{}{}
\makeatother
\IfFileExists{footnotehyper.sty}{\usepackage{footnotehyper}}{\usepackage{footnote}}
\makesavenoteenv{longtable}
\usepackage{graphicx}
\makeatletter
\newsavebox\pandoc@box
\newcommand*\pandocbounded[1]{
  \sbox\pandoc@box{#1}%
  \Gscale@div\@tempa{\textheight}{\dimexpr\ht\pandoc@box+\dp\pandoc@box\relax}%
  \Gscale@div\@tempb{\linewidth}{\wd\pandoc@box}%
  \ifdim\@tempb\p@<\@tempa\p@\let\@tempa\@tempb\fi
  \ifdim\@tempa\p@<\p@\scalebox{\@tempa}{\usebox\pandoc@box}%
  \else\usebox{\pandoc@box}%
  \fi%
}
\def\fps@figure{htbp}
\makeatother
\NewDocumentCommand\citeproctext{}{}

\makeatletter
 \let\@cite@ofmt\@firstofone
 \def\@biblabel#1{}
 \def\@cite#1#2{{#1\if@tempswa , #2\fi}}
\makeatother
\newlength{\cslhangindent}
\newlength{\csllabelwidth}
\newenvironment{CSLReferences}[2] 
 {\begin{list}{}{%
  \setlength{\itemindent}{0pt}
  \setlength{\leftmargin}{0pt}
  \setlength{\parsep}{0pt}
  \ifodd #1
   \setlength{\leftmargin}{\cslhangindent}
   \setlength{\itemindent}{-1\cslhangindent}
  \fi
  \setlength{\itemsep}{#2\baselineskip}}}
 {\end{list}}
\usepackage{calc}

\providecommand{\tightlist}{%
  \setlength{\itemsep}{0pt}\setlength{\parskip}{0pt}}
\usepackage{todonotes}
\usepackage{pdfcomment}
\usepackage{amsmath,cleveref,caption}
\usepackage{bbm}
\makeatletter
\def\fps@figure{htb}
\makeatother
\usepackage{subfig}
\usepackage{bookmark}
\IfFileExists{xurl.sty}{\usepackage{xurl}}{} 
\hypersetup{
  pdftitle={Don't Drop the Singletons: Efficient Inference for Pairwise Experiments with Independent Attrition},
  pdfauthor={Simon He, Patrick W. Schmidt},
  hidelinks,
  pdfcreator={LaTeX via pandoc}}

\title{Don't Drop the Singletons: Efficient Inference for Pairwise Experiments with Independent Attrition}
\author{Simon He\ss, Patrick W. Schmidt}
\date{\today}

\begin{document}
\maketitle
\begin{abstract}
Pairwise randomization can yield substantial efficiency gains in experiments. Yet methodological guidance cautions against pairwise randomization, especially in settings with attrition, partly because common practices for estimation (i.e., pair fixed effects) imply discarding data from incomplete pairs thus exacerbating data loss from attrition. This practice of dropping incomplete pairs reduces statistical power of tests as well as precision of estimates, in paired experiments, compared to designs with less finely stratified treatment assignment. We argue that this concern is misplaced if attrition is independent of treatment status and potential outcomes, and that these issues follow from an inefficient use of the data that remains post-attrition. First, we show how, by using a specific permutation test, it is possible to use all observed units for inference (complete pairs and incomplete pairs where one unit attrits) while still exploiting the pairwise randomization design structure. The test procedure we suggest provides exact size control under the sharp null. Second, we study an optimally weighted estimator that efficiently combines within-pair and across-pair comparisons. Finally, we show that combining these two insights yields a test procedure that dominates the two commonly used inference methods (a paired \(t\)-test and the two-sample \(t\)-test) in power, for any level of attrition. Usefully for applied researchers, we show that the efficient procedure can be implemented via a weighted fixed effects regression, straightforward in standard software. In sum, our results provide researchers with practical tools for conducting experiments with pairwise randomization without sacrificing observations or statistical power when facing independent attrition.
\end{abstract}

\section{Introduction}\label{sec:introduction}

Experiments with pairwise randomization, i.e., randomly assigning treatment to one unit in blocks of two units each, can offer substantial efficiency gains over less finely stratified randomization, particularly when sample sizes are limited.
As surveyed in Bai, Shaikh, and Tabord-Meehan (forthcoming), pairwise randomization designs have unique advantages for estimation precision and test power. Imai, King, and Nall (2009) demonstrate for cluster-randomized experiments that pair matching can dramatically increase statistical power, arguing that ``from the perspective of bias, efficiency, power, robustness or research costs, and in large or small samples, pairing should be used in cluster-randomized experiments whenever feasible.'' Bruhn and McKenzie (2009) provide simulation evidence supporting this claim, finding that pairwise randomization can outperform alternative randomization strategies in achieving balance in potential outcomes when covariates with good predictive power for outcomes exist (and the researchers have a single outcome to focus on).
More recently, Bai (2022) establishes formal optimality results, showing that pairwise randomization designs emerge as optimal among a large set of randomization schemes.
These theoretical and empirical results suggest pairwise designs should be widely adopted where efficiency matters.

Despite these advantages, methodological guidance has cautioned against pairwise randomization designs when attrition is anticipated, for a number of reasons.
The widely-used evaluation handbook by Glennerster and Takavarasha (2013, 159) advises researchers facing attrition risk to ``use strata that have at least four units rather than pairwise randomization.''\footnote{Not to be confused with Athey and Imbens (2017)'s argument that groups larger than pairs can provide within-stratum variance estimators, that may improve variance estimation and thus inference, even without attrition. This is a different argument than the one about attrition, and we do not focus on it in this paper.} Similarly, Donner and Klar (2000) characterize the need to drop singletons (incomplete pairs) as a fundamental weakness of the approach (Bai et al. 2024 cover this debate in more detail). Likewise, in a blogpost, McKenzie (2022) identifies attrition as ``main reason for being cautious'' regarding pairwise randomization, also noting that when units drop out, the pairwise randomization designs are usually analyzed in a way that discards not only the unit that attrits but also its paired counterpart, the orphaned singletons, thus increasing effective attrition rates. Our paper specifically addresses this concern by suggesting and discussing a test method that does not discard these singletons.

The core of our paper is based on the observation that if attrition occurs independently of treatment assignment, the canonical test procedures for pairwise designs (such as the paired \(t\)-test) unnecessarily discard useful information.\footnote{Where attrition depends on treatment and potential outcomes, the problem is a different one and identification itself is at stake (Bai et al. 2024)}
We argue that this can be overcome; i.e., that it is possible to conduct valid and powerful inference while fully exploiting the pairwise design structure and using all observed units, including the orphaned singletons in pairs with attrition. Specifically, we show that pairwise designs analyzed with our methods dominate standard inference approaches. Whether they also dominate the alternatives of designs with larger strata is a related but different question that remains open and that we discuss in the conclusion and through a short simulation study.

Conveniently, our efficient procedure can be estimated via a single weighted fixed-effects regression. So researchers who anticipate exogenous attrition can retain the efficiency benefits of pairing at essentially no implementation cost, rather than falling back to a study design with coarser stratification.

We build on a literature on the econometrics of finely stratified experiments that has seen some new progress in the past years.
Bai, Romano, and Shaikh (2022) study inference under pairwise randomization \emph{without} attrition: they show that both the two-sample and the paired \(t\)-tests are asymptotically conservative, propose an adjusted variance estimator that restores asymptotic exactness, and show that the within-pair permutation test is finite-sample valid under the sharp null and, when studentized by the adjusted variance, asymptotically exact under the weak null.
Our point of departure is an observation in Bai (2022, Online Appendix C.3) :
the difference-in-means, computed without dropping singletons, remains consistent for the ATE under two conditions: attrition is independent of treatment assignment (conditional on baseline covariates if included), and
attrition is independent of the individual treatment effect.
No variance, test, or distinction between complete pairs and singletons is developed there.
We build on this observation and develop it into an inference procedure that separates the complete-pair and singleton components, derives their variance-minimizing combination, and provides exact randomization-based tests.
Bai et al. (2024) study pairwise randomization \emph{with} possibly endogenous attrition, deriving the estimands recovered by the difference-in-means estimator both when singletons are dropped and when they are retained. They find limited support for the practice of dropping singletons, showing that under heterogeneous treatment effects, the difference in means when singletons are dropped generally identifies a weighted average of conditional treatment effects rather than the ATE.
Their focus is identification, whereas we hold the estimand fixed by assuming independent attrition and optimize the inference procedure: we show how retained singletons should be combined with complete pairs, quantify the resulting power gain, and provide exact tests.
Relatedly, Fukumoto (2022) investigates implications of the two common ways of dealing with outcome-correlated attrition: either dropping or retaining singletons in the presence of attrition. These two testing approaches (below we refer to those as the paired \(t\)-test, often implemented via a regression with pair fixed effects, and two-sample \(t\)-test, often implemented via a regression without fixed effects) have been widely used in practice when analyzing pairwise randomized experiments with attrition.

Our main contribution is to show that the choice between the two standard approaches, paired \(t\)-test and two-sample \(t\)-test, is a false dichotomy, and to provide a combined alternative that produces valid inference, has optimal power in a clearly defined setting, and is easy to implement. Both standard approaches waste information in different ways: the paired \(t\)-test discards all singletons, thus losing information from singletons, while the two-sample \(t\)-test ignores the pairwise design structure, thus wasting information about the design. By combining randomization inference with an optimally weighted estimator that efficiently combines information from complete pairs and singletons, we can make full use of all information for the estimand, and by using randomization inference, we can conduct valid inference that respects the pairwise design structure, providing exact size control under the sharp null for any attrition rate, while dominating standard approaches in power.\footnote{A related literature in biostatistics on incomplete paired data dates back to Ekbohm (1976). Closest to us is Amro and Pauly (2017), who construct a permutation test also combining a complete-pairs test statistic with an unpaired-singletons statistic.
  Our approach differs in four respects:
  First, our framework allows for optimal combination. We combine two unbiased estimators with weights derived to minimize the variance of the combined estimator, which maximizes power of the resulting test for our setting. Second, we combine estimators directly rather than pre-standardized test statistics. Third, in our setting exactness follows from the treatment-assignment mechanism in an RCT, without distributional assumptions. Fourth, and building on the first three points, we derive closed-form power and a dominance result.}

In detail, our paper's contribution is threefold:
First, we propose a permutation-based randomization inference procedure that respects the pairwise design structure while using all observed units after attrition: complete pairs and singletons. This procedure provides exact size control under the sharp null hypothesis for any attrition rate when attrition is independent of treatment assignment and potential outcomes.
Second, we introduce an optimally weighted estimator that combines information from within-pair differences (from complete pairs) and between-unit comparisons (from singletons). This estimator is efficient in a clearly defined setting and, under constant effects, yields a test that dominates both the paired \(t\)-test (which discards observations) and the two-sample \(t\)-test (which ignores pairing) in power.
Third, we show that the optimally weighted estimator can be computed via a weighted regression with appropriate observation weights, enabling straightforward implementation in standard statistical software. This works via weighted OLS, by regressing the outcome on treatment status and \emph{pair fixed effects only for complete pairs}, while including singletons with a lower per-observation weight that reflects their contribution to the variance of the estimator.

When facing attrition, researchers often present results across multiple methods with varying identifying assumptions, from strongest to weakest. Our optimally weighted randomization inference procedure assumes attrition is independent of treatment assignment and, under this assumption, uses all observed units optimally.
Analyses based on our procedure should thus be presented next to more conservative approaches that relax independence, such as the endogenous-attrition identification analysis of Bai et al. (2024) or bounding methods for non-random attrition, with our estimate serving as the point obtained under the strongest assumption about attrition against which those weaker-assumption analyses can be compared.

The remainder of the paper is organized as follows. \Cref{sec:setting} presents our formal setting, describes the four inference procedures we compare, and introduces our optimally weighted estimator. This section also derives the asymptotic power of each procedure as a function of attrition rate and matching quality.
\Cref{sec:results} presents our main results: we illustrate the power functions across different scenarios, establish exact size control under the sharp null, demonstrate the dominance of our optimally weighted procedure, and extend our approach to weak-null testing via studentized randomization inference.
\Cref{sec:conclusion} concludes with a discussion of practical implementation and directions for future research. All proofs are collected in the appendix.

\section{Setting}\label{sec:setting}

\subsection{Model and Data Generating Process}\label{model-and-data-generating-process}

We consider experiments with \(N = 2n\) units organized into \(n\) pairs. Each pair is drawn independently from a common population of pairs, i.e., we study this question through the lens of a superpopulation model with sampling uncertainty, where pairs are sampled.
Our sampling model follows the pairs-as-primitive superpopulation perspective of Fukumoto (2022), who, following Imai (2008) and Imbens and Rubin (2015) (ch.~10), draws the sampled pairs from a superpopulation of pairs; we differ in drawing pairs i.i.d. from a bivariate distribution \(F\) rather than from a large finite population of pairs, and in restricting attention to attrition that is independent of assignment.\footnote{Our framework also differs from the superpopulation model for pairs of Bai, Romano, and Shaikh (2022), in which units are sampled i.i.d. and pairs are formed on observed covariates, so that within-pair dependence is endogenous and heterogeneous across pairs; we instead sample pairs i.i.d. from a fixed joint distribution \(F\) with a common within-pair correlation, trading covariate-adaptive generality for closed-form power expressions and finite-sample statements.
  Two remarks on this choice.
  First, it is made for ease of exposition of our main argument. With pairs as i.i.d. draws, match quality reduces to the single parameter \(r\), the optimal weights and all power functions below have closed forms, and the comparison across inference methods can be read directly off the resulting expressions.
  Second, the choice is not what drives our results and recommendation for applied researchers. The exact size control of the randomization tests (Proposition 6) is a finite-sample statement conditional on potential outcomes and the attrition pattern; it uses only the within-pair randomization and holds verbatim whether pairs are sampled from \(F\), units are sampled i.i.d. and matched on covariates as in Bai, Romano, and Shaikh (2022), or the sample is a fixed population as in Chaisemartin and Ramirez-Cuellar (2024). Likewise, the dominance of the optimally weighted procedure rests on inverse-variance weighting of two unbiased, uncorrelated component estimators, of which the mean within-pair difference weights \((1,0)\) and the difference-in-means weights \((1-q,\,q)\) are special cases; this logic is not specific to our sampling model. What is framework-specific is the closed-form characterization of weights and power in terms of \((q, r)\), and the validity of unstudentized weak-null inference, which we delimit in \Cref{weak-null-discussion}.
  So, a reader who prefers a different framework can retain the substantive conclusion: under attrition that is independent of treatment assignment, the default choice between discarding singletons and ignoring the pairing is dominated.} We index pairs by \(p=1, ..., n\) and units within pairs by \(g\in\{1,2\}\). Specifically, for a pair, the untreated potential outcomes of its two units, \((Y_{p,1}(0), Y_{p,2}(0))\), are an i.i.d. draw from a bivariate distribution \(F\). The two coordinates of \(F\) are exchangeable, i.e.~swapping which unit is labeled 1 vs.~2 leaves the distribution unchanged, so both members share a common marginal distribution. Every ``for large \(n\)'' statement below refers to \(n \to \infty\) pairs drawn from \(F\).\footnote{Asymptotic normality follows from the Lindeberg--Feller central limit theorem for triangular arrays (Vaart 1998, Proposition 2.27).} This framework is chosen to resemble a setting in which population pairs pre-exist (e.g., neighbors) or are formed through some matching procedure (e.g., pairing on Mahalanobis distance of baseline covariates), which is encoded in the dependence structure of \(F\).

We denote the marginal variance of the untreated potential outcome by \(\sigma^2_{Y_0} = \text{Var}(Y_{p,g}(0))\), and require it to be finite. The within-pair correlation
\[r = \text{Corr}(Y_{p,1}(0), Y_{p,2}(0)) \in [0,1]\]
is a functional of \(F\) and hence a population parameter. It is typically unobserved and can be understood as a measure of match quality.
Higher values of \(r\) indicate better matching quality. In the limit case, \(r = 0\), pairing provides no efficiency gain over complete randomization; conversely as \(r \to 1\), paired units have identical control potential outcomes.\footnote{If treatment effects are heterogeneous rather than a constant shift, \(r\) remains well-defined as a property of the untreated-outcome but its estimation from post-treatment data becomes challenging.}

For tractability, we first consider constant treatment effects. We define \(Y_{p,g}(1) := Y_{p,g}(0) + \tau\), so the treated potential outcome is derived from the drawn untreated outcome rather than sampled separately. Our power calculations use this constant-shift specification. Our results extend straightforwardly to a location-shift family, \(Y_{p,g}(1) \overset{d}{=} Y_{p,g}(0) + \tau\), which permits heterogeneous effects while fixing the average treatment effect at \(\tau\); we return to discussing unconstrained heterogeneity in the second half of the paper.

Two further layers of randomness exist on top of the pair sampling: treatment assignment and attrition.
First, treatment assignment: within every pair, one unit is assigned to treatment and the other to control by an independent coin flip, decided independently of the drawn outcomes and applied to all pairs before any attrition.
Treatment status is denoted by the indicator \(D_{p,g}\in\{0,1\}\); for notational convenience we write \(T\) (treated, \(D=1\)) and \(C\) (control, \(D=0\)) as subscripts on group means and variances.
Second, attrition: exactly a fraction \(q\) of outcomes is removed deterministically, yielding exactly \(m = n(1-q)^2\) complete pairs, \(k_1 = nq(1-q)\) observed treated singletons, and \(k_0 = nq(1-q)\) observed control singletons.\footnote{This deterministic specification keeps the observed counts fixed and preserves the feature that matters, balance of attrition by treatment status; under random unit-level attrition the counts become random but converge to the same values and the power formulas are unchanged in the limit. Further, we ignore integer-rounding of \(n(1-q)^2\), \(nq(1-q)\), and \(nq^2\) throughout, which affects the results only at \(o(1)\).}
We call pair \(p\) \emph{complete} if both units are observed, and a \emph{singleton} if only one unit is observed.

All results below are based on this framework, with its three independent sources of randomness: the draw of pairs from \(F\), the within-pair assignment, and attrition. The asymptotic power results (Propositions 1--5) are driven by the sampling of pairs from \(F\). Finite-sample exact size control (Proposition 6) follows by conditioning on the realized draw and attrition pattern and using only the within-pair assignment randomization; since this holds for every draw, it also holds unconditionally over the draw.

\subsection{Considered Inference Methods}\label{considered-inference-methods}

We consider four inference procedures, all based on two-sided tests at level \(\alpha\).
The first two methods are standard approaches commonly used in practice, each with known limitations under attrition.
The latter two methods are randomization inference procedures that respect the pairwise design structure while using all observed units.

\textbf{Paired \(t\)-test.} Using only the \(m = n(1-q)^2\) complete pairs, the estimator is the mean within-pair difference
\[\hat{\tau}_{\text{pair}} = \frac{1}{m}\sum_{p:\,\text{complete}} (Y_{p,T} - Y_{p,C}),\]
with sampling variance
\[\text{SE}_{\text{pair}}^2 = \frac{2\sigma^2_{Y_0}(1-r)}{n(1-q)^2},\]
(derived in the proof of Proposition 1). Inference uses the \(t\)-statistic \(\hat{\tau}_{\text{pair}}/\widehat{\text{SE}}_{\text{pair}}\) against a \(t\) reference with \(m-1\) degrees of freedom. Under attrition this discards all singleton observations. Empirically, economists typically implement this test in a fixed-effects regression: \(\hat{\tau}_{\text{pair}}\) is the coefficient on \(D_{p,g}\) in the OLS regression \(Y_{p,g} = \beta D_{p,g} + \alpha_p + u_{p,g}\) with pair fixed effects \(\alpha_p\), where the pair fixed effects drop the singletons.

\textbf{Two-sample \(t\)-test.} Using all observed units regardless of pair status, the estimator used for this test is the difference in means \[\hat{\tau} = \bar{Y}_T - \bar{Y}_C.\] The two-sample \(t\)-test computes its standard error under the (incorrect) assumption of complete randomization,
\[\text{SE}_{\text{assumed}}^2 = \frac{2\sigma^2_{Y_0}}{n(1-q)},\]
and refers \(\hat{\tau}/\widehat{\text{SE}}_{\text{assumed}}\) to a \(t\) (large-\(n\): normal) distribution. Empirically, economists typically implement this test in a plain OLS regression: \(\hat{\tau}\) is the coefficient on \(D_{p,g}\) in the OLS regression of \(Y_{p,g}\) on \(D_{p,g}\) with a constant.

The variance used for inference in this test, \(\text{SE}_{\text{assumed}}^2\), is not the true sampling variance of \(\hat{\tau}\). Under the pairwise design, the true sampling variance of \(\hat{\tau}\) over the draw from \(F\), the within-pair assignment, and attrition is
\begin{equation}
\text{SE}_\tau^2 = \frac{2\sigma^2_{Y_0}[1 - r + qr]}{n(1-q)}, \label{eq:se-tau}
\end{equation}
derived in the proof of Proposition 3. Since \(\text{SE}_\tau^2 \le \text{SE}_{\text{assumed}}^2\), the two-sample \(t\)-test is typically conservative: it ignores the design-based efficiency gain from pairing. The two-sample \(t\)-test and the difference-in-means randomization test below share the estimator \(\hat{\tau}\) and the true sampling variance \(\text{SE}_\tau^2\); they differ only in the reference distribution used for inference. The former uses a plug-in normal with the mis-specified \(\text{SE}_{\text{assumed}}\), the latter uses the correct permutation distribution.\footnote{A third variant is possible: refer \(\hat{\tau}\) to a normal using a consistent estimate of the correct variance \(\text{SE}_\tau^2\) (via \(\hat{\sigma}^2_{Y_0}\) and \(\hat{r}\)) rather than \(\text{SE}_{\text{assumed}}\). In the setting without attrition, inference based on a consistent estimate of the correct variance is the main proposal of Bai, Romano, and Shaikh (2022); our variant is in the same spirit, though we do not establish a formal correspondence between the two.}

\textbf{Randomization inference (RI) using the difference in means as test statistic.}
Using all observed units, we compute the observed difference in means \(\hat\tau = \bar{Y}_T - \bar{Y}_C\).
We obtain the null distribution by repeatedly permuting\footnote{For complete pairs this amounts to randomly exchanging the treatment status between the observations.
  For singleton observations, since attrition is by assumption balanced by treatment status (exactly \(nq(1-q)\) observed treated and the same number of control units from singleton pairs), each permutation randomly reassigns their treatment labels while maintaining this balance, exactly like complete randomization with fixed group sizes.} treatment assignment within each pair and recalculating the test statistic for each permutation.
The two-sided \(p\)-value is the proportion of permutations, \(\pi\), yielding \(|\tau^{(\pi)}| \geq |\hat\tau|\).

This procedure respects the pairwise design structure through the permutation scheme while using all observed units, thus avoiding the limitations of the paired \(t\)-test and two-sample \(t\)-test under attrition. The permutation scheme is design-aware: complete-pair members are swapped within pair, singletons are permuted unrestricted (or more precisely: within pair but with an unobserved counterpart). The test statistic, however, weights every observation equally in \(\bar{Y}_T - \bar{Y}_C\), and so does not combine the complete-pair and singleton information according to their differing precision. The optimally weighted procedure below also addresses this remaining inefficiency through weighting.

\textbf{Randomization inference using an optimally weighted estimator as test statistic.}
The above RI procedure uses a simple difference in means as the test statistic, treating all observed units identically.
However, under the pairwise design with attrition, we can partition our observed sample into two distinct subsets with different precision:
complete pairs (for which we observe both treatment and control outcomes and can thus difference out pair-level heterogeneity) and
singletons (for which we observe only one unit and must estimate treatment effects from between-pair comparisons).
This partitioning suggests the existence of an \emph{optimally weighted estimator} that combines both according to their relative precision, which we derive below. Its sampling variance \(\text{SE}_{\text{oracle}}^2\) is given in Proposition 4.
For inference, we use this optimally weighted estimator as the test statistic in the randomization inference procedure described above.
The next subsection derives this optimally weighted estimator in detail.

\subsubsection{Optimally weighted estimator}\label{optimally-weighted-estimator}

The optimally weighted estimator is a linear combination of the two unbiased estimators for the average treatment effect \(\tau\), so, as algebraic building blocks, we define separate estimators for the two data partitions: complete pairs and singletons.

For complete pairs,
\[\hat{\tau}_{\text{pair}} = \frac{1}{m}\sum_{p:\, \text{complete}} (Y_{p,T} - Y_{p,C})\]
where \(Y_{p,T}\) and \(Y_{p,C}\) denote the observed outcomes of the treated and control member of pair \(p\), respectively, \(m= n(1-q)^2\) is the number of complete pairs, and the within-pair difference eliminates pair-specific heterogeneity.

For singletons, \[\hat{\tau}_{\text{single}} = \bar{Y}_{T,\text{single}} - \bar{Y}_{C,\text{single}}\] uses between-unit comparisons (retaining full \(\sigma^2_{Y_0}\) variance).

Both estimators are unbiased and efficient for the average treatment effect \(\tau\) within their respective data partitions.
The optimally weighted treatment effect estimator is then:
\[\hat\tau_{\text{oracle}} = w^* \cdot \hat{\tau}_{\text{pair}} + (1-w^*) \cdot \hat{\tau}_{\text{single}}\]
where \(w^*\) is chosen to minimize the variance of \(\hat\tau_{\text{oracle}}\) via standard inverse-variance weighting:
\[w^* = \frac{V_{\text{single}}}{V_{\text{pair}} + V_{\text{single}}}\]
where \(V_{\text{pair}} = 2\sigma^2_{Y_0}(1-r)/[n(1-q)^2]\) and \(V_{\text{single}} = 2\sigma^2_{Y_0}/[nq(1-q)]\) are the variances of the two component estimators.

We call this the \emph{oracle} estimator because \(w^*\) depends on population parameters (useful for power calculations); the \emph{feasible} estimator \(\hat\tau_{\text{feasible}}\) replaces \(w^*\) with an estimate \(\hat{w}^*\) and is defined below.

\paragraph{Oracle optimally weighted estimator}\label{oracle-optimally-weighted-estimator}

The oracle optimal weight simplifies to
\[w^* = \frac{(1-q)}{(1-q) + q(1-r)}.\]
By efficiently combining two efficient estimators, the oracle estimator achieves the minimum possible variance among all unbiased linear combinations of \(\hat{\tau}_{\text{pair}}\) and \(\hat{\tau}_{\text{single}}\), and therefore provides an efficiency bound for this setting.\footnote{We conjecture this is the efficiency bound in a stronger sense, i.e., that no regular estimator using the observed units achieves smaller asymptotic variance, since the two components together exhaust the observed-data information about \(\tau\). We do not pursue this here.}

\paragraph{Feasible optimally weighted estimator}\label{feasible-optimally-weighted-estimator}

In practice, the population variances and correlation are unknown and must be estimated from the data. The optimal weight \(w^*\) depends on the attrition rate \(q\) and the within-pair correlation \(r\). The attrition rate can be directly observed as \(\hat{q} = 1 - n_{\text{observed}}/N\), where \(n_{\text{observed}} = 2m + k_1 + k_0\) is the number of non-attrited units and \(N = 2n\). Estimating the within-pair correlation \(\hat{r}\) requires more care, and we discuss two approaches.

When baseline outcomes \(Y_{p,g}^{\text{baseline}}\) are available, the within-pair correlation can be estimated directly from pre-treatment data. Compute the within-pair variance \(\hat{\sigma}_{WP,\text{baseline}}^2 = \frac{1}{n-1} \sum_{p=1}^n (Y_{p,1}^{\text{baseline}} - Y_{p,2}^{\text{baseline}})^2\) and the marginal variance \(\hat{\sigma}_{Y,\text{baseline}}^2 = \frac{1}{2n-1} \sum_{p=1}^{n}\sum_{g \in \{1,2\}} (Y_{p,g}^{\text{baseline}} - \bar{Y}^{\text{baseline}})^2\), then estimate \(\hat{r}^{\text{baseline}} = 1 - \hat{\sigma}_{WP,\text{baseline}}^2/(2\hat{\sigma}_{Y,\text{baseline}}^2)\). This approach directly measures the matching quality achieved by the pairing procedure and avoids contamination from post-treatment heterogeneity. It assumes that baseline correlation predicts the correlation in control potential outcomes, which is plausible when outcomes are relatively stable and matching was based on baseline predictors.

We conjecture that even absent baseline data, a pre-specified educated guess for \(r\) (for instance, based on pilot studies or similar experiments) is likely to improve power over standard practice of using a paired \(t\)-test. The paired \(t\)-test, which discards all singleton observations, implicitly corresponds to setting \(\hat{r} = 1\) (optimal only under perfect matching), an assumption that is almost never justified in practice. Our optimally weighted approach nests the paired \(t\)-test as the limiting case when \(r \to 1\), but allows for the more realistic scenario of imperfect matching, thereby recovering information from singletons that would otherwise be discarded.

Without baseline data or educated guesses, the within-pair correlation could be estimated from post-treatment outcomes:

\begin{itemize}
\tightlist
\item
  Within-pair variance: \(\hat{\sigma}_{WP}^2 = \frac{1}{m-1} \sum_{p:\, \text{complete}} (\Delta_p - \hat{\tau}_{\text{pair}})^2\), where \(\Delta_p = Y_{p,T} - Y_{p,C}\) is the treated-minus-control within-pair difference and \(\hat{\tau}_{\text{pair}} = \frac{1}{m}\sum_{p:\,\text{complete}}\Delta_p\) their mean.
\item
  Marginal variance: \(\hat{\sigma}_{Y_0}^2 = \frac{1}{n_0-1} \sum_{(p,g): D_{p,g}=0} (Y_{p,g} - \bar{Y}_0)^2\) where \(n_0 = m + k_0\) is the number of observed control units (control members of complete pairs plus control singletons), with \(n_0 = n(1-q)\).
\item
  Estimated within-pair correlation: \(\hat{r} = 1 - \frac{\hat{\sigma}_{WP}^2}{2\hat{\sigma}_{Y_0}^2}\)
\item
  Observed attrition rate: \(\hat{q} = 1 - n_{\text{observed}}/N\)
\end{itemize}

It is worth noting that for the purpose of sharp hypothesis testing via randomization inference, this approach requires no assumptions beyond those already maintained for the sharp null hypothesis \(H_0: Y_{p,g}(1) = Y_{p,g}(0)\), since under this null the within-pair variance estimates the within-pair variance in control outcomes exactly, with no contribution from treatment effect heterogeneity. The baseline data approach, when available, provides more stable estimates using all pairs and avoids potential complications from heterogeneous treatment effects when the estimator is used for purposes beyond testing (such as estimation or power calculations).

When using the feasible estimator with sample-estimated components as a test statistic in randomization inference, all variance estimates and weights should be recomputed for each permutation. Specifically, for each permuted treatment assignment \(D^{(\pi)}\), compute \(\hat{\sigma}_{Y_0}^2(\pi)\) using observations with \(D^{(\pi)} = 0\), derive \(\hat{r}(\pi)\) and \(\hat{w}^*(\pi)\) accordingly, and use \(\hat{w}^*(\pi)\) to compute the permuted test statistic.
The within-pair variance is recomputed as \(\hat{\sigma}_{WP}^2(\pi) = \frac{1}{m-1}\sum_{p:\,\text{complete}}(\Delta_p^{(\pi)} - \hat{\tau}_{\text{pair}}(\pi))^2\) for each permutation, since \(\Delta_p^{(\pi)} = Y_{p,T(\pi)} - Y_{p,C(\pi)}\) and its mean both change sign for pairs whose assignment is flipped. Under the sharp null this recompute is immaterial (it coincides with the fixed member-label version at \(\tau = 0\)) while under the alternative it is what keeps the estimate unbiased.
This procedure ensures that the permutation distribution properly reflects uncertainty in both the treatment effect and the nuisance parameters, preserving exact size control under the sharp null.

\subsubsection{Estimation}\label{estimation}

This paper focuses on inference, but the optimally weighted estimator can also be used for estimation and is efficient. Conveniently, the estimator can also be computed via a weighted regression with appropriately chosen weights for singletons vs complete pairs.

\textbf{Lemma (Optimally Weighted Estimator via Weighted Regression).} The oracle optimally weighted estimator can be obtained as the treatment coefficient from the weighted OLS regression:
\[Y_{p,g} = \beta D_{p,g} + \alpha_0\mathbbm{1}\{p \text{ is singleton}\} +  \alpha_p\mathbbm{1}\{p \text{ is complete pair}\} + \varepsilon_{p,g}\]
with observation-level weights proportional to the precision of each design component:
\[\tilde{w}_{\text{pair}} \propto \frac{1}{1-r}, \quad \tilde{w}_{\text{single}} \propto 1\]

In plain language, this regression specification uses all observed units, but includes pair fixed effects only for complete pairs, while singletons are absorbed into a separate intercept and complete pairs are weighted more heavily, reflecting their higher precision in estimating the treatment effect.
The resulting coefficient on \(D_{p,g}\) is equivalent to the optimally weighted estimator \(\hat\tau_{\text{oracle}}\).

\emph{Proof.} See appendix \cref{proof:weighted-regression}.

The observation weights implement inverse-variance weighting based on design-specific information content. Paired observations identify \(\tau\) through within-pair differences with variance proportional to \((1-r)\), while singleton observations identify \(\tau\) through unpaired comparisons with variance proportional to \(1\). The weights optimally balance the efficiency gain from pairing: when \(r\) is high, paired observations are heavily upweighted; when \(r \to 0\), weights converge to equal weighting.

\subsection{Asymptotic power as a function of attrition and matching quality}\label{asymptotic-power-as-a-function-of-attrition-and-matching-quality}

Below, we derive the asymptotic power of each of the four inference procedures under constant treatment effects \(\tau\) as a function of attrition rate \(q\) and matching quality \(r\).\footnote{All power statements are along local alternatives \(\tau_n = h/\sqrt{n}\) for fixed \(h\): the formulas give the limiting rejection probability with \(\tau\) replaced by \(\tau_n\). This is needed because at a fixed \(\tau \neq 0\) the permutation reference distribution is inflated by terms of order \(\tau^2/\sigma^2_{Y_0}\) (e.g., for pairs \(E[\Delta_p^2] = 2\sigma^2_{Y_0}(1-r) + \tau^2\)), so the formulas evaluated at fixed \(\tau\) overstate power by a relative error of that order; along \(\tau_n = h/\sqrt{n}\) this term is \(o(1)\). Propositions 1 and 2 use the same convention for comparability. For the calibrations in \Cref{sec:results}, \(\tau^2/\sigma^2_{Y_0} \approx 1\%\), so the fixed-\(\tau\) reading is accurate to that order.}
First, the pairwise design with independent attrition efficiently uses information in complete pairs, but loses information from attrited units.
Second, the two-sample \(t\)-test uses all observed units but has two shortcomings.
The difference in means used in this test fails to exploit the pairwise design structure, by treating observations from complete pairs the same as singletons, thus inefficiently combining available information. The standard error estimate used in this test also fails to account for the pairwise design structure, leading to incorrect and typically conservative standard errors.
Third, the randomization inference procedure using the difference in means as test statistic remedies the second shortcoming by respecting the pairwise design structure through the permutation scheme, while still using all observed units. However, like the two-sample \(t\)-test, it treats all observed units identically in the construction of the test statistic and thus does not remedy the first shortcoming of inefficiently combining information from pairs and singletons.

Finally, the randomization inference procedure using the optimally weighted estimator as test statistic remedies both shortcomings, efficiently combining information from complete pairs and singletons according to their relative precision, while respecting the pairwise design structure through the permutation scheme.

\subsubsection{\texorpdfstring{Paired \(t\)-test}{Paired t-test}}\label{paired-t-test}

\textbf{Proposition 1.} Under constant treatment effects along local alternatives \(\tau_n = h/\sqrt{n}\), the asymptotic power of the paired \(t\)-test at significance level \(\alpha\) is approximately:
\[\text{Power}_{\text{Paired}} = \Phi\left(\frac{\tau\sqrt{n(1-q)^2}}{\sqrt{2\sigma^2_{Y_0}(1-r)}} - z_{\alpha/2}\right) + \Phi\left(-\frac{\tau\sqrt{n(1-q)^2}}{\sqrt{2\sigma^2_{Y_0}(1-r)}} - z_{\alpha/2}\right)\]
where \(\Phi\) is the standard normal CDF and \(z_{\alpha/2} = \Phi^{-1}(1-\alpha/2)\).

\emph{Proof.} See appendix \cref{proof:power-paired-t-test}.

\subsubsection{\texorpdfstring{Two-sample \(t\)-test (difference in means)}{Two-sample t-test (difference in means)}}\label{two-sample-t-test-difference-in-means}

\textbf{Proposition 2.} Under constant treatment effects, \(\tau\), along local alternatives and pairwise randomization, the asymptotic power of the two-sample \(t\)-test (which ignores pairing) at significance level \(\alpha\) is approximately:
\[\text{Power}_{\text{Mean Diff}} = \Phi\left(\lambda(\delta - z_{\alpha/2})\right) + \Phi\left(\lambda(-\delta - z_{\alpha/2})\right)\]
where
\[\delta = \frac{\tau\sqrt{n(1-q)}}{\sqrt{2\sigma^2_{Y_0}}}\]
is the non-centrality parameter under the \(t\)-test's assumption of complete randomization, and
\[\lambda = \sqrt{\frac{\text{SE}_{\text{assumed}}^2}{\text{SE}_{\tau}^2}}\]
is the variance inflation factor accounting for the mismatch between the \(t\)-test's assumption (complete randomization) and the actual design (pairwise randomization). Here:
\[\text{SE}_{\text{assumed}}^2 = \frac{2\sigma^2_{Y_0}}{n(1-q)}\]
\[\text{SE}_{\tau}^2 = \frac{2\sigma^2_{Y_0}[1 - r + qr]}{n(1-q)} \text{(as given by \eqref{eq:se-tau})}\]

\emph{Proof.} See appendix \cref{proof:power-2-sample-t-test}.

\subsubsection{Randomization Inference (RI) with Difference in Means as Test Statistic}\label{randomization-inference-ri-with-difference-in-means-as-test-statistic}

\textbf{Proposition 3.} Under constant treatment effects along local alternatives \(\tau_n = h/\sqrt{n}\), the asymptotic power of the randomization inference test at significance level \(\alpha\) is approximately:
\[\text{Power}_{\text{RI, Mean Diff}} = \Phi\left(\frac{\tau}{\text{SE}_{\tau}} - z_{\alpha/2}\right) + \Phi\left(-\frac{\tau}{\text{SE}_{\tau}} - z_{\alpha/2}\right)\]
where \(\text{SE}_{\tau}^2\) is the same as in Proposition 2.

\emph{Proof.} See appendix \cref{proof:power-ri-diff-means}.

\subsubsection{Randomization Inference with an Optimally Weighted Estimator as Test Statistic}\label{randomization-inference-with-an-optimally-weighted-estimator-as-test-statistic}

\paragraph{Oracle optimally weighted RI}\label{oracle-optimally-weighted-ri}

\textbf{Proposition 4.} Under constant treatment effects along local alternatives \(\tau_n = h/\sqrt{n}\) and known population parameters (variances and \(r\)), the asymptotic power of the oracle optimally weighted RI test at significance level \(\alpha\) is approximately:
\[\text{Power}_{\text{oracle}} = \Phi\left(\frac{\tau}{\text{SE}_{\text{oracle}}} - z_{\alpha/2}\right) + \Phi\left(-\frac{\tau}{\text{SE}_{\text{oracle}}} - z_{\alpha/2}\right)\]
where

\[\text{SE}_{\text{oracle}}^2 = \frac{2\sigma^2_{Y_0}(1-r)}{n(1-q)(1 - rq)}\]

\emph{Proof.} See appendix \cref{proof:power-ri-optimal-oracle}.

\textbf{Analytic (oracle) test.} Proposition 4 yields an analytic test alongside the randomization test: refer \(\hat{\tau}_{\text{oracle}}/\text{SE}_{\text{oracle}}\) to the standard normal. It shares the estimator and sampling variance of the optimally weighted randomization test, and by the CLT established in the proof of Proposition 4 has the same asymptotic power \(\text{Power}_{\text{oracle}}\); the two differ only in reference distribution. The randomization test is preferred because it retains exact size under the sharp null (Proposition 6), which the normal approximation does not. Replacing \(\text{SE}_{\text{oracle}}\) with a consistent estimate inflates the variance by \(O(n^{-3/2})\) (Proposition 5), so the feasible analytic test is asymptotically equivalent. The power curves in \Cref{sec:results} are computed from this closed-form expression.

\textbf{Proposition 5 (Asymptotic equivalence of feasible and oracle procedures).} Under constant treatment effects along local alternatives \(\tau_n = h/\sqrt{n}\) and assuming \(F\) has finite fourth moments, the power of the feasible optimally weighted RI test approaches that of the oracle procedure as \(n \to \infty\):
\[\text{Power}_{\text{feasible}} \to \text{Power}_{\text{oracle}} = \Phi\left(\frac{\tau}{\text{SE}_{\text{oracle}}} - z_{\alpha/2}\right) + \Phi\left(-\frac{\tau}{\text{SE}_{\text{oracle}}} - z_{\alpha/2}\right)\]

More precisely, the feasible estimator satisfies:
\[\text{SE}_{\text{feasible}}^2 = \text{SE}_{\text{oracle}}^2 + O(n^{-3/2})\]

\emph{Proof.} See appendix \cref{proof:power-ri-optimal-feasible}.

\section{Results and Discussion}\label{sec:results}

\subsection{Illustration}\label{illustration}

\begin{figure}
\subfloat[Weak matching benefit\label{fig:power-plots-1}]{\includegraphics[width=0.33\linewidth]{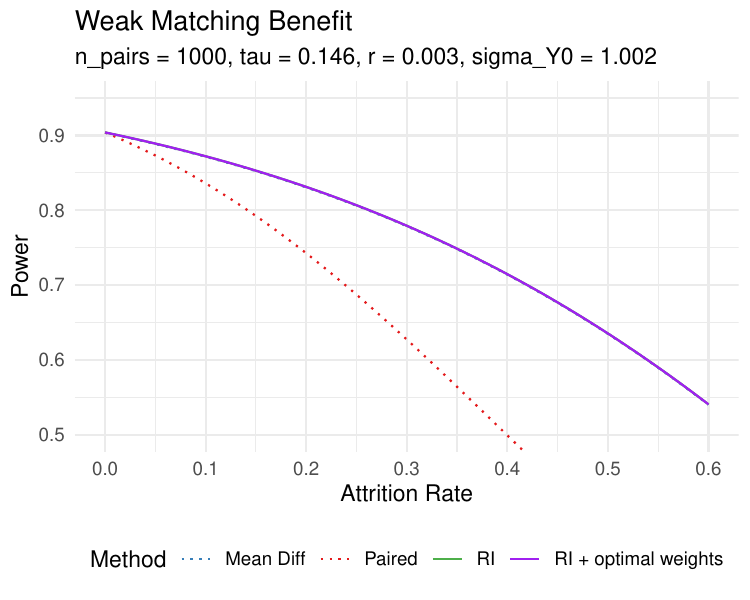} }\subfloat[Moderate matching benefit\label{fig:power-plots-2}]{\includegraphics[width=0.33\linewidth]{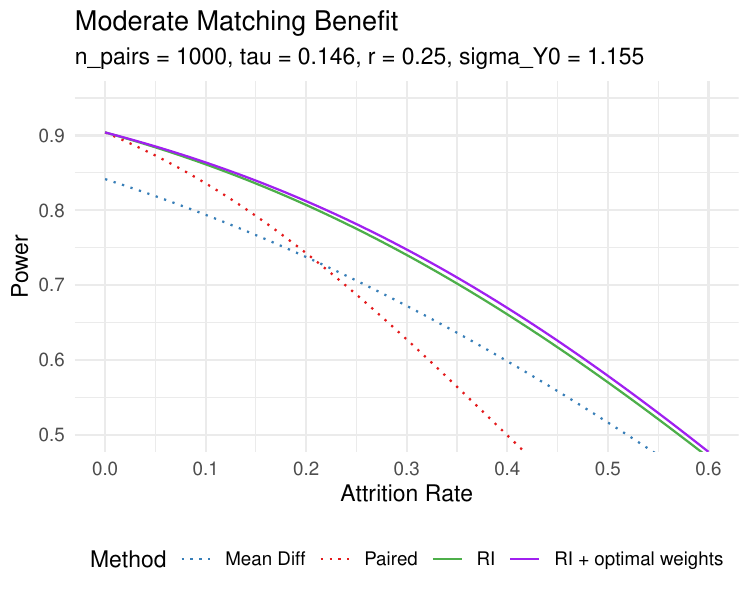} }\subfloat[Strong matching benefit\label{fig:power-plots-3}]{\includegraphics[width=0.33\linewidth]{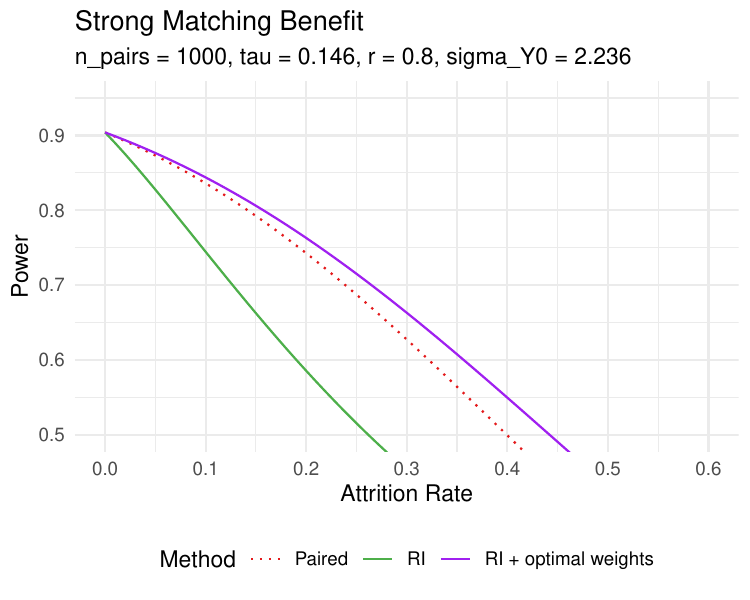} }\caption{Power as a function of attrition}\label{fig:power-plots}
\end{figure}

In figure \ref{fig:power-plots}, we illustrate the theoretical power results for the four inference methods across varying attrition rates and matching quality. As a useful benchmark case, we chose a parametrization where, in the absence of attrition (\(q=0\)), the paired \(t\)-test achieves 90\% power. This corresponds to an effect size of \(\tau = 0.146\) and \(n = 1000\) pairs. To enable direct comparison across matching quality levels, we hold the baseline noise variance constant while allowing the marginal variance \(\sigma^2_{Y_0}\) to vary with \(r\) (higher matching quality implies higher correlation in potential outcomes, thus higher marginal variance for fixed noise). For the chosen parametrization, we consider three scenarios of matching quality:

\textbf{Weak matching (\(r = 0.003\)):} All methods start off with similar power at \(q=0\), as matching provides minimal efficiency gain, and pairwise matching effectively resembles complete randomization. As attrition increases, the paired \(t\)-test loses power quickly due to attrition of complete pairs, while the two-sample \(t\)-test maintains power by using all observed units. Both RI methods perform similarly to the two-sample \(t\)-test.

\textbf{Moderate matching (\(r = 0.25\)):} The paired \(t\)-test starts off at par with the two RI methods at \(q=0\), but loses power quickly as attrition increases and the paired \(t\)-test suffers from the high attrition of complete pairs. At high attrition rates, the two-sample \(t\)-test outperforms the paired \(t\)-test, because it uses all observed units. Both RI methods dominate the two standard approaches across all attrition rates.

\textbf{Strong matching (\(r = 0.80\)):} The paired \(t\)-test starts off with substantially higher power at \(q=0\), leveraging the strong matching. However, as attrition increases, the paired \(t\)-test again loses power due to attrition of complete pairs. The two-sample \(t\)-test's power is below 50\%. RI with difference in means also performs poorly as attrition increases, as it fails to efficiently combine information from pairs and singletons. RI with the optimally weighted estimator dominates all other methods across all attrition rates, leveraging both the strong matching and using all observed units.

It is worth noting that in practice, we may not know the exact matching quality in advance. However, the optimally weighted RI procedure is robust to this uncertainty, as it estimates the optimal weights from the data, thus adapting to the actual matching quality.

\subsection{RI is well sized also with attrition}\label{ri-is-well-sized-also-with-attrition}

As randomization inference directly leverages the random assignment mechanism, it provides exact finite-sample size control under the sharp null hypothesis of no treatment effect, even in the presence of attrition, and without a need for large-sample approximations or distributional assumptions.

\textbf{Proposition 6 (Exact size control).} Under the sharp null hypothesis \(H_0: Y_{p,g}(1) = Y_{p,g}(0)\) for all \((p,g)\) and the assumption that attrition is independent of treatment assignment
conditional on the pair's full vector of potential outcomes \(\mathbf{Y}_p := (Y_{p,1}(0), Y_{p,1}(1), Y_{p,2}(0), Y_{p,2}(1))\), i.e.~\(A_{p,g} \perp D_{p,g} \mid \mathbf{Y}_p\), both RI procedures (standard and optimally weighted) reject with probability at most \(\alpha\) (exactly \(\alpha\) under a randomized tie-break) for any finite sample size \(n\) and attrition rate \(q\).

\emph{Proof.} See appendix \cref{proof:exact-size-control}.

\textbf{Remark (holds also with larger strata).} This argument does not rely on the pair structure and holds for any finely-stratified design with assignment-independent attrition. The estimator, weights, and power expressions above, by contrast, are specific to pairs.

\textbf{Remark (Implementation with feasible weights).} In implementing the feasible weighted RI procedure, the weights \(\hat{w}^*(\pi)\) must be recomputed for each permutation \(\pi\) using the variance estimates corresponding to that permutation's treatment assignment. Specifically, for each permutation, compute \(\hat{\sigma}_{Y_0}^2(\pi)\) using observations labeled as control under that permutation, derive \(\hat{r}(\pi)\) and \(\hat{w}^*(\pi)\) accordingly, and compute the test statistic using these permutation-specific weights. This recomputation ensures exact size control under the sharp null, as the permutation distribution properly accounts for the joint randomization distribution of both the treatment effect estimator and the nuisance parameter estimates. Holding weights fixed across permutations (computed only from the observed treatment assignment) would break exactness because the control group variance estimate \(\hat{\sigma}_{Y_0}^2\) is not permutation-invariant.

\textbf{Remark on asymptotic vs exact results:} Unlike the power results (Propositions 1-5), which rely on asymptotic normality via the Lindeberg--Feller CLT, Proposition 6 provides exact finite-sample size control under the sharp null without any distributional assumptions beyond the independence of attrition from treatment assignment.

\subsection{Weak-null inference}\label{weak-null-discussion}

The derivations above are exact under the sharp null \(H_0: Y_{p,g}(1) = Y_{p,g}(0)\) for all \((p,g)\).
We now consider the weak null \(H_0: E[Y_{p,g}(1) - Y_{p,g}(0)] = 0\), which permits arbitrary effect heterogeneity.
Under the weak null, permutation-based inference is not automatically valid: the permutation distribution of a test statistic need not match its true sampling distribution once treatment effects vary across units (Chung and Romano 2013; Ding and Dasgupta 2018; Wu and Ding 2021). The standard remedy is studentization, which restores asymptotic validity under the weak null while typically preserving exact size under the sharp null. We show below that in our setting this remedy is not always needed.

\subsubsection{Without studentization}\label{without-studentization}

The reason is that both test statistics (the within-pair difference as well as differences between singletons), as well as their {[}weighted{]} combination, are balanced in treatment. Consider the singleton component first. Treated and control singletons are equal in number, so the difference in means is a two-sample comparison with equal group sizes. For such a comparison the permutation variance and the true sampling variance coincide when the two groups are equal in size, even if their outcome distributions differ (Chung and Romano 2013, Example 2.1). The weak null is therefore tested at the correct level without studentization. The pair component of the test statistic is a within-pair sign flip of the differences \(\Delta_p = Y_{p,T} - Y_{p,C}\). Its permutation variance is \(E[\Delta_p^2]\) and its true variance is \(E[\Delta_p^2] - (E\Delta_p)^2\); these agree exactly when \(E\Delta_p = 0\), which is the weak null. The sign flip is balanced by construction, so no tuning is needed. The full estimator combines the two components, which use disjoint partitions of the data with independent assignments. The two permutation pieces are therefore independent and each matches its sampling counterpart under the null, so their weighted sum does as well.

Balance is what these arguments rely on.
For singletons, we assumed exactly equal numbers of treated and control singletons for simplicity. But also more generally, when attrition is independent of treatment, the two counts are equal in expectation and their ratio tends to one, so the equal-size condition holds in the limit and the singleton component is asymptotically valid regardless of the realized counts. For pairs, symmetry holds by construction.\footnote{This does not contradict the studentization requirement of Bai, Romano, and Shaikh (2022). In their framework, units are sampled i.i.d. and then paired on covariates, and under the weak null the within-pair sign-flip permutation distribution of the mean pair difference has limiting variance exceeding the estimator's sampling variance by \(\tfrac{1}{2}E[(E[Y_i(1)-Y_i(0)\mid X_i])^2]\): matching makes the two members of a pair share the same conditional average effect, which the estimator averages over all \(2n\) sampled units while the permutation treats each pair's difference as a single draw. In our model the pair itself is the i.i.d. draw, so the permutation variance of the pair component, \(E[\Delta_p^2]\), coincides with its sampling variance \(\mathrm{Var}(\Delta_p)\) whenever \(E[\Delta_p]=0\), whatever the within-pair dependence of the effects. The cost is that our model cannot express effect heterogeneity systematically related to the pairing index across pairs; where that is a concern, the studentized procedures below apply.}

\subsubsection{With studentization}\label{with-studentization}

If one prefers not to rely on this balance, expects attrition to differ across arms while remaining independent, or operates in a different sampling framework,
the estimator can instead be studentized by a consistent standard error, which restores weak-null validity for any group sizes while keeping exact size under the sharp null.\footnote{In general, consistency for the sampling variance is not sufficient for permutation tests. The variance estimator used for studentization must be consistent for the variance of the randomization distribution, which may be different (Bai, Romano, and Shaikh 2022, Remark 3.16); in the framework that we used here, the two limits coincide under the weak null, so this distinction has no bite.} This requires a variance estimator appropriate to the pairwise design. Consistent alternatives are discussed for different cases in the literature: Bai, Romano, and Shaikh (2022), adapted to attrition by Bai et al. (2024), construct one by grouping adjacent pairs, and Chaisemartin and Ramirez-Cuellar (2024) discusses when clustering at the strata level is appropriate. A reader who wants full robustness to heterogeneous effects and unbalanced attrition can adopt either estimator, studentize each component, and test the weak null in the same permutation framework.

\subsection{Optimally weighted RI outperforms the other methods for all parameter values}\label{optimally-weighted-ri-outperforms-the-other-methods-for-all-parameter-values}

\textbf{Proposition 7 (Dominance of optimally weighted RI).} For any \(n, q, \sigma^2_{Y_0}, r\) with \(q \in (0,1)\) and \(r \in [0,1)\), the oracle optimally weighted RI procedure has power greater than or equal to both the paired \(t\)-test and the two-sample \(t\)-test.

\emph{Proof.} See appendix \cref{proof:dominance-optimal-ri}.

Figure \ref{fig:advantage-plots} shows the advantage of the optimally weighted RI procedure over the two standard methods across varying attrition rates and matching quality. The advantage is measured as the percentage reduction in required sample size to achieve the same power as the optimally weighted RI procedure, when using the paired \(t\)-test or the two-sample \(t\)-test instead.

\begin{figure}
\subfloat[RI with optimal weights vs. Paired t-test\label{fig:advantage-plots-1}]{\includegraphics[width=0.33\linewidth]{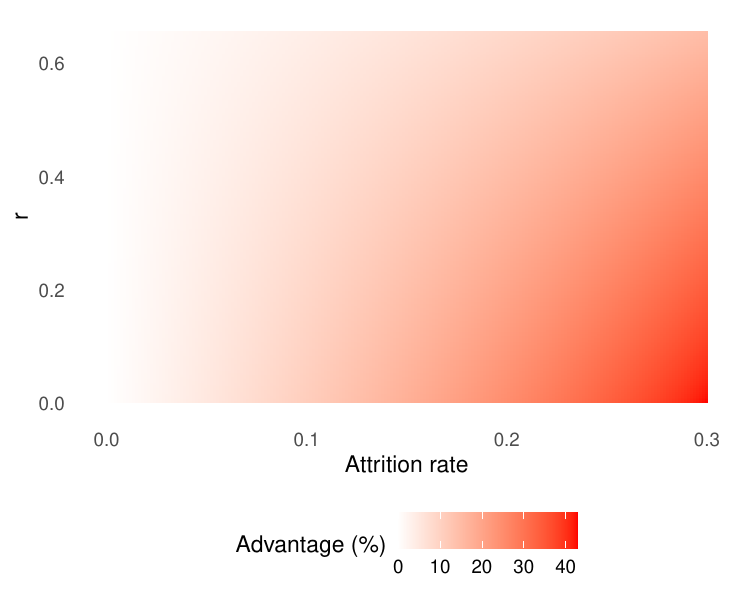} }\subfloat[RI with optimal weights vs. Two-sample t-test\label{fig:advantage-plots-2}]{\includegraphics[width=0.33\linewidth]{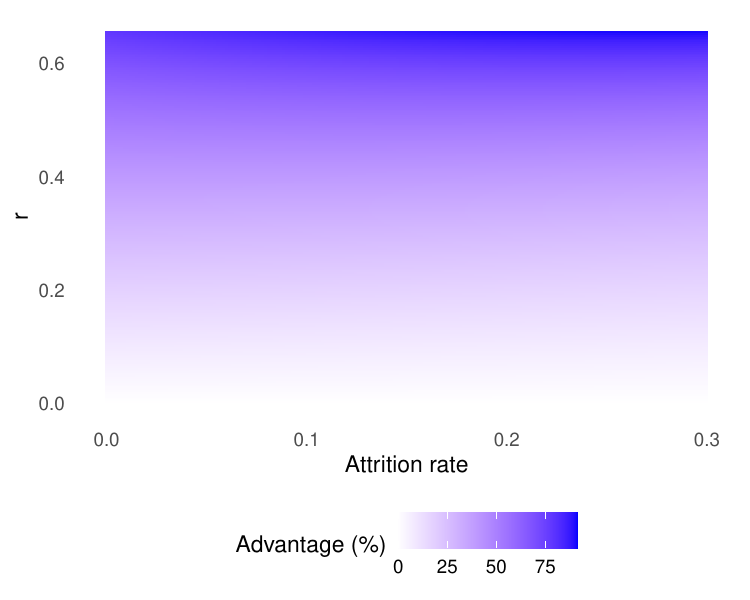} }\subfloat[RI versus best of both\label{fig:advantage-plots-3}]{\includegraphics[width=0.33\linewidth]{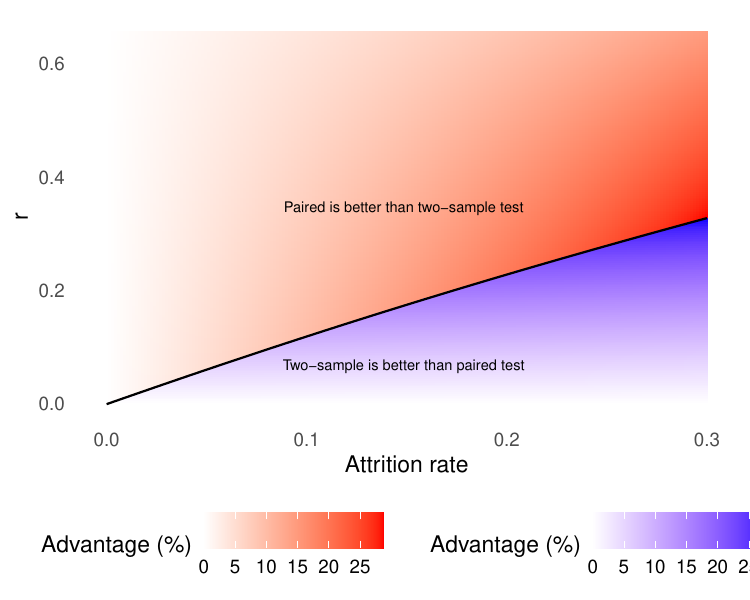} }\caption{Advantage of RI with optimal weights (\%-reduction in required sample size to achieve target power) over standard methods}\label{fig:advantage-plots}
\end{figure}

\section{Simulation to compare pair designs with our method to minimally larger strata}\label{simulation-to-compare-pair-designs-with-our-method-to-minimally-larger-strata}

\subsection{Simulation: Paired vs.~Larger Strata Designs}\label{simulation-paired-vs.-larger-strata-designs}

The methodological guidance discussed above commonly recommends using larger strata (e.g., strata of size four) rather than pairs when substantial attrition is anticipated.
Our theoretical results establish that randomization inference with optimal weights dominates standard approaches within pairwise designs, but leave open whether pairwise designs analyzed with our methods can compete with larger strata designs analyzed with standard methods.
We investigate this through simulation.

Strata of size four offer advantages in a specific regime: attrition rates high enough that many pairs become singletons, but not so high that four-unit strata lose all within-stratum variation.
However, this advantage depends critically on the trade-off between matching quality and effective sample size.
When tight matching gains are substantial, pairwise designs may still dominate even with higher effective attrition.
When gains from matching are modest, larger strata may be better---but in those cases, the benefits of efficiently including singletons are also higher, thus improving the performance of our optimally weighted inference.

\subsection{Simulation design}\label{simulation-design}

We simulate data from the theoretical model of \Cref{sec:setting}, with additional structure to enable comparison across designs. We generate \(N = 2n\) units with a baseline covariate \(X_{p,g} \sim N(0, \sigma_x^2)\) and idiosyncratic noise \(\varepsilon_{p,g} \sim N(0, \sigma_\varepsilon^2)\), yielding control potential outcomes \(Y_{p,g}(0) = X_{p,g} + \varepsilon_{p,g}\). Treatment effects are constant: \(Y_{p,g}(1) = Y_{p,g}(0) + \tau\).

To ensure fair comparison, both designs are constructed from the same realized data by sorting units on \(X_{p,g}\). The pairwise design assigns consecutive units to pairs; the strata-of-4 design assigns consecutive groups of four to strata.
This is to mimic the decision a researcher would make when forming matched pairs or strata based on a single pre-treatment covariate that is a decent predictor of outcomes.
Treatment is randomized independently within each pair or stratum.
Attrition occurs independently at rate \(q\) per unit, independent of the employed design.
This construction isolates the pure design comparison: pairs achieve tighter matching (higher within-pair correlation in \(Y_{p,g}(0)\)), while pairs have a higher probability of losing all within-stratum variation due to attrition.

This setup imposes stronger assumptions than our theoretical analysis requires, e.g., normal distributions, a specific matching procedure, and serves only to demonstrate existence of parameter configurations where our proposed method's efficiency gains offset the attrition disadvantage of pairwise designs.

The chosen setup is also such that stratifying at a higher level has virtually no disadvantages absent attrition. We chose the parametrization such that the within strata variance of potential outcomes is almost identical irrespective of whether strata of 2 or 4 are formed. This is reflected by the fact that in the simulations below, all designs achieve similar power, when there is no attrition. We do this to show that the advantage of our method does not only exist when larger strata are disadvantaged anyways.

\subsection{Results}\label{results}

\begin{table}[!h]
\centering
\caption{\label{tab:sim_results}Simulated power comparison: Paired design with optimal RI vs. strata-of-4 design.}
\centering
\begin{tabular}[t]{rrrr}
\toprule
\multicolumn{1}{c}{ } & \multicolumn{3}{c}{Power} \\
\cmidrule(l{3pt}r{3pt}){2-4}
\multicolumn{1}{c}{ } & \multicolumn{2}{c}{Pairs} & \multicolumn{1}{c}{Strata-of-4} \\
\cmidrule(l{3pt}r{3pt}){2-3} \cmidrule(l{3pt}r{3pt}){4-4}
Attrition Rate & Pair FE & RI Optimal & Strata FE\\
\midrule
0.00 & 0.892 & 0.895 & 0.891\\
0.05 & 0.864 & 0.874 & 0.853\\
0.10 & 0.813 & 0.839 & 0.834\\
0.15 & 0.759 & 0.802 & 0.789\\
0.20 & 0.711 & 0.779 & 0.759\\
\addlinespace
0.25 & 0.662 & 0.735 & 0.716\\
0.30 & 0.607 & 0.713 & 0.689\\
0.35 & 0.560 & 0.671 & 0.645\\
0.40 & 0.503 & 0.628 & 0.570\\
\bottomrule
\end{tabular}
\end{table}

Table \ref{tab:sim_results} presents simulated power at \(\alpha = 0.05\) for three methods
(pairwise randomization design with paired \(t\)-test,
strata-of-4 design with strata-fixed effects,
pairwise design with optimally weighted randomization inference) across attrition rates. Parameters are \(n = 50\) pairs, \(\tau = 0.146\), \(\sigma_x = 0.130\), \(\sigma_\varepsilon = 0.224\) (chosen to yield approximately 90\% power for the pairwise design without attrition). With no attrition, all methods achieve almost exactly the nominal power of 90\%.
As attrition increases to 15\%, pair fixed effects drops to 76\% power while the strata-of-4 design maintains 79\%.
However, optimally weighted RI on pairs achieves 80\%, surpassing both.
At high levels of 30\% attrition, the paired \(t\)-test drops to 61\%, reflecting that in expectation only \((1-0.3)^2 = 49\%\) of pairs remain complete.
The strata-of-4 design with fixed effects maintains 69\%, still marginally below the optimally weighted RI on pairs at 71\%.

These results demonstrate that efficiency gains from tight matching in pairs, when properly exploited through optimal weighting of complete pairs and singletons, can still dominate the effective sample size advantage of larger strata in terms of power.
Researchers facing anticipated attrition need not abandon pairwise designs if they adopt the inference methods we propose. And researchers who are constrained to pairwise designs can still achieve substantial power gains by employing our optimally weighted randomization inference approach.

\section{Conclusion and discussion}\label{sec:conclusion}

Some practitioner advice cautions against pairwise randomization when attrition is expected. One argument in that direction is that loss of complete pairs when inference is based on a paired t-test will reduce power. We argue that the case against pairwise randomization rests, under ``ignorable'' attrition, on the choice of a suboptimal inference method rather than on a defect of the design itself. When attrition is independent of treatment, i.e., ``ignorable'', tests that use every observed unit (complete pairs as well as orphaned singletons) while still exploiting the paired structure, can be constructed and retain exact size under the sharp null at any attrition rate. Weighting complete pairs and singletons by their precision yields a test that dominates both the paired \(t\)-test and the two-sample \(t\)-test in power under standard assumptions. The procedure we propose is straightforwardly implemented as a weighted fixed-effects regression, in standard software.

Our analysis assumes homoscedasticity in the sense that pairs are exchangeable draws from a single distribution \(F\), so that the within-pair correlation \(r\) and marginal variance \(\sigma^2_{Y_0}\) are common across pairs. Applied researchers may not want to make this assumption. Relaxing exchangeability, by allowing pair-specific correlations \(r_p\) and variances \(\sigma^2_{Y_0,p}\), as would arise when match quality varies across the sample, would make the optimal weight pair-specific. We conjecture that a generalization of the dominance result (Proposition 7) holds under such heterogeneity, but leave this extension to future work.

A fruitful extension could be to generalize our approach to more experimental designs, such as stratified randomization with larger strata, and compare the efficiency gains across designs \emph{and} methods (e.g., inference with the corrected variance estimator of Bugni, Canay, and Shaikh (2018)).
We believe the result of these comparisons is not obvious. Without attrition, block size has no first-order effect on asymptotic efficiency of the unadjusted estimator for RCTs with a fixed treated fraction (Bai et al. 2025); attrition is precisely what makes the design comparison interesting.
When attrition becomes an issue, designs with larger strata can continue to exploit within-stratum variation from partially complete strata. So they might outperform pairs if the gains from using within strata variation outweigh the gain coming from tighter matching in pairs. This is context-specific. Formalizing this trade-off into generalizable rules could be valuable at the design stage of an RCT, but we leave it for future research.

\section*{References}\label{references}
\addcontentsline{toc}{section}{References}

\protect\phantomsection\label{refs}

\clearpage
\appendix

\section{Appendix A: Proofs}\label{appendix-a-proofs}

\subsection{\texorpdfstring{Power of Paired \(t\)-test}{Power of Paired t-test}}\label{proof:power-paired-t-test}

With attrition rate \(q\), the number of complete pairs is \(m=n(1-q)^2\).
Condition on the attrition pattern; since attrition is independent of the draw from \(F\), the \(m\) complete pairs are an i.i.d. sample from \(F\). For each complete pair \(p\), under constant treatment effects the within-pair difference is:
\[\Delta_p = Y_{p,T} - Y_{p,C} = \tau + (Y_{p,T}(0) - Y_{p,C}(0))\]
where \(Y_{p,T}(0) - Y_{p,C}(0)\) captures differences in control potential outcomes within the pair.

Let \(S_p \in \{-1,+1\}\) denote the within-pair coin flip, independent of the draw and i.i.d. across pairs with \(E[S_p] = 0\), so that \(Y_{p,T}(0) - Y_{p,C}(0) = S_p(Y_{p,1}(0) - Y_{p,2}(0))\). Then \(\Delta_p - \tau = S_p(Y_{p,1}(0) - Y_{p,2}(0))\), and since \(S_p^2 = 1\) and \(S_p\) is independent of the draw,
\[\text{Var}(\Delta_p) = E[(Y_{p,1}(0) - Y_{p,2}(0))^2].\]
By exchangeability the two members share a common marginal mean, so the member difference has mean zero and
\[E[(Y_{p,1}(0) - Y_{p,2}(0))^2] = \text{Var}(Y_{p,1}(0)) + \text{Var}(Y_{p,2}(0)) - 2\text{Cov}(Y_{p,1}(0), Y_{p,2}(0)) = 2\sigma^2_{Y_0}(1-r).\]

Under the alternative hypothesis with effect \(\tau\), we have \(E[\bar{\Delta}] = \tau\) and \(\text{Var}(\bar{\Delta}) = 2\sigma^2_{Y_0}(1-r)/[n(1-q)^2]\).
For large \(n(1-q)^2\), by the Lindeberg--Feller CLT (Vaart 1998, Proposition 2.27), the test statistic \(t = \bar{\Delta}/\text{SE}(\bar{\Delta})\) under the alternative is approximately distributed as \(N(\delta, 1)\) where
\(\delta = \tau\sqrt{n(1-q)^2}/\sqrt{2\sigma^2_{Y_0}(1-r)}\).

The two-sided test rejects when \(|t| > z_{\alpha/2}\). Under the alternative:
\[\text{Power} = \Phi(\delta - z_{\alpha/2}) + \Phi(-\delta - z_{\alpha/2}) = \Phi\left(\frac{\tau\sqrt{n(1-q)^2}}{\sqrt{2\sigma^2_{Y_0}(1-r)}} - z_{\alpha/2}\right) + \Phi\left(-\frac{\tau\sqrt{n(1-q)^2}}{\sqrt{2\sigma^2_{Y_0}(1-r)}} - z_{\alpha/2}\right) \quad \square\]

\subsection{\texorpdfstring{Power of Two-sample \(t\)-test}{Power of Two-sample t-test}}\label{proof:power-2-sample-t-test}

After attrition, we observe \(2n(1-q)\) units total, with \(n(1-q)\) in each treatment arm. The two-sample \(t\)-test statistic is:
\[t = \frac{\bar{Y}_T - \bar{Y}_C}{\widehat{\text{SE}}(\bar{Y}_T - \bar{Y}_C)}\]

\textbf{The \(t\)-test assumes complete randomization.} Under this assumption with common variance \(\sigma^2_{Y_0} = \text{Var}(Y_{p,g}(0))\):
\[\text{SE}_{\text{assumed}}(\bar{Y}_T - \bar{Y}_C) = \sqrt{\frac{2\sigma^2_{Y_0}}{n(1-q)}}\]

Under the alternative hypothesis with effect \(\tau\), if the data had actually come from complete randomization, the \(t\)-test would have non-centrality parameter:
\[\delta = \frac{\tau}{\text{SE}_{\text{assumed}}} = \frac{\tau\sqrt{n(1-q)}}{\sqrt{2\sigma^2_{Y_0}}}\]

\textbf{However, the data come from pairwise randomization, not complete randomization.} Under pairwise randomization, the true sampling variance of \(\bar{Y}_T - \bar{Y}_C\) is \(\text{SE}_\tau^2\) from \eqref{eq:se-tau},
\[\text{SE}_{\tau}^2 = \frac{2\sigma^2_{Y_0}[1-r + qr]}{n(1-q)},\]
whose derivation is given in the proof of Proposition 3 below.

This is smaller than \(\text{SE}_{\text{assumed}}^2\) because pairwise randomization constrains treatment assignment within pairs, reducing variance.

\textbf{Distribution of the test statistic.} Under the alternative, the numerator \(\bar{Y}_T - \bar{Y}_C\) has mean \(\tau\) and variance \(\text{SE}_{\tau}^2\), while the denominator \(\widehat{\text{SE}}_{\text{assumed}}\) converges to \(\text{SE}_{\text{assumed}}\). Therefore:
\[t = \frac{\bar{Y}_T - \bar{Y}_C}{\widehat{\text{SE}}_{\text{assumed}}} \sim N\left(\frac{\tau}{\text{SE}_{\text{assumed}}}, \left(\frac{\text{SE}_{\tau}}{\text{SE}_{\text{assumed}}}\right)^2\right) = N\left(\delta, \frac{1}{\lambda^2}\right)\]

where the variance inflation factor is:
\[\lambda = \sqrt{\frac{\text{SE}_{\text{assumed}}^2}{\text{SE}_{\tau}^2}}.\]

This asymptotic distribution follows from the Lindeberg--Feller CLT (Vaart 1998, Proposition 2.27) applied to the sample means in the numerator.

\textbf{Power calculation.} The two-sided test rejects when \(|t| > z_{\alpha/2}\). To find this probability, standardize the test statistic:
\[\lambda(t - \delta) \sim N(0, 1)\]

Then:
\[P(t > z_{\alpha/2}) = P(\lambda(t - \delta) > \lambda(z_{\alpha/2} - \delta)) = \Phi(\lambda\delta - \lambda z_{\alpha/2})\]

Similarly:
\[P(t < -z_{\alpha/2}) = P(\lambda(t - \delta) < \lambda(-z_{\alpha/2} - \delta)) = \Phi(-\lambda\delta - \lambda z_{\alpha/2})\]

Therefore:
\[\text{Power} = \Phi(\lambda\delta - \lambda z_{\alpha/2}) + \Phi(-\lambda\delta - \lambda z_{\alpha/2}) = \Phi(\lambda(\delta - z_{\alpha/2})) + \Phi(\lambda(-\delta - z_{\alpha/2})) \quad \square\]

\textbf{Remark:} When \(q = 0\) (no attrition), \(\lambda = 1/\sqrt{1-r}\). When \(r\) is high (strong matching), pairwise randomization substantially reduces variance compared to complete randomization, and \(\lambda > 1\), increasing power even though the \(t\)-test ignores pairing.

\subsection{Power of Randomization Inference}\label{proof:power-ri-diff-means}

The RI test statistic is the difference in means \(\hat{\tau} = \bar{Y}_T - \bar{Y}_C\) using all observed units. After attrition, we have \(n(1-q)^2\) complete pairs, \(nq(1-q)\) treatment singletons, and \(nq(1-q)\) control singletons.

\textbf{Step 1: Decompose the test statistic.} The treatment and control means can be written as weighted averages:
\[\bar{Y}_T = (1-q) \cdot \bar{Y}_T^{\text{pairs}} + q \cdot \bar{Y}_{T,\text{single}}\]
\[\bar{Y}_C = (1-q) \cdot \bar{Y}_C^{\text{pairs}} + q \cdot \bar{Y}_{C,\text{single}}\]
where the superscripts denote means within complete pairs and singletons respectively. Therefore:
\[\hat{\tau} = (1-q)(\bar{Y}_T^{\text{pairs}} - \bar{Y}_C^{\text{pairs}}) + q(\bar{Y}_{T,\text{single}} - \bar{Y}_{C,\text{single}})\]

\textbf{Step 2: Variance from complete pairs.}

Condition on the attrition pattern; since attrition is independent of the draw from \(F\), the \(m = n(1-q)^2\) complete pairs are an i.i.d. sample from \(F\). For each complete pair \(p\), let \(S_p \in \{-1, +1\}\) denote the within-pair coin flip determining which member is treated, independent of the draw and i.i.d. across pairs with \(E[S_p] = 0\). Centered at the treatment effect \(\tau\), pair \(p\) contributes
\[\frac{S_p}{m}\big(Y_{p,1}(0) - Y_{p,2}(0)\big)\]
to \(\bar{Y}_T^{\text{pairs}} - \bar{Y}_C^{\text{pairs}}\). These contributions are i.i.d. across pairs (i.i.d. draws from \(F\), independent coin flips), so the variance over both sources of randomness is
\[\text{Var}\!\left[\frac{S_p}{m}\big(Y_{p,1}(0) - Y_{p,2}(0)\big)\right] = \frac{1}{m^2}\, E\big[(Y_{p,1}(0) - Y_{p,2}(0))^2\big],\]
using \(E[S_p] = 0\) and \(S_p^2 = 1\). By exchangeability the two members share a common marginal mean, so the member difference has mean zero and
\[E\big[(Y_{p,1}(0) - Y_{p,2}(0))^2\big] = \text{Var}\big(Y_{p,1}(0) - Y_{p,2}(0)\big) = \sigma^2_{Y_0} + \sigma^2_{Y_0} - 2r\sigma^2_{Y_0} = 2\sigma^2_{Y_0}(1-r).\]
Summing over the \(m\) i.i.d. pairs,
\[\text{Var}\big[\bar{Y}_T^{\text{pairs}} - \bar{Y}_C^{\text{pairs}}\big] = \frac{2\sigma^2_{Y_0}(1-r)}{n(1-q)^2}.\]

\textbf{Step 3: Variance from singletons.} Conditional on the attrition pattern there are exactly \(k_1 = k_0 = nq(1-q)\) treated and control singletons, each an i.i.d. draw from the marginal of \(F\) and independent of all other units. Conditional on the observed counts \(k_1 = k_0 = nq(1-q)\), the independent within-pair coin flips make all \(\binom{k_1+k_0}{k_1}\) treated/control labelings of the singletons equally likely, so permuting singleton labels with fixed group sizes is a valid conditional randomization test. The two singleton means are then independent, each with variance \(\sigma^2_{Y_0}/(nq(1-q))\), giving
\[\text{Var}\big[\bar{Y}_{T,\text{single}} - \bar{Y}_{C,\text{single}}\big] = \frac{\sigma^2_{Y_0}}{k_1} + \frac{\sigma^2_{Y_0}}{k_0} = \frac{2\sigma^2_{Y_0}}{nq(1-q)},\]
using \(k_1 = k_0 = nq(1-q)\).

\textbf{Step 4: Combine variance components.} Since pair draws, coin flips, and singleton draws are mutually independent:
\[\text{Var}(\hat{\tau}) = (1-q)^2 \cdot \frac{2\sigma^2_{Y_0}(1-r)}{n(1-q)^2} + q^2 \cdot \frac{2\sigma^2_{Y_0}}{nq(1-q)}\]

Simplifying:
\[\text{Var}(\hat{\tau}) = \frac{2\sigma^2_{Y_0}(1-r)}{n} + \frac{2q\sigma^2_{Y_0}}{n(1-q)}\]

Combining the first two terms over a common denominator:
\begin{align*}
&= \frac{2\sigma^2_{Y_0}(1-r)(1-q) + 2q\sigma^2_{Y_0}}{n(1-q)}  \\
&= \frac{2\sigma^2_{Y_0}[(1-r)(1-q) + q]}{n(1-q)}  \\
&= \frac{2\sigma^2_{Y_0}[1 - r + qr]}{n(1-q)} 
\end{align*}

\[\text{SE}_{\tau}^2 = \frac{2\sigma^2_{Y_0}[1 - r + qr]}{n(1-q)},\] which is the \(\text{SE}_\tau^2\) of \eqref{eq:se-tau}.

\textbf{Step 5: Normal approximation.} Write \(\hat{\tau} - \tau\) as the sum of \(m\) i.i.d. pair contributions and \(k_1 + k_0\) i.i.d. singleton contributions. The two blocks have different within-block laws (pair terms scale with \(2\sigma^2_{Y_0}(1-r)\), singleton terms with \(2\sigma^2_{Y_0}\)), so \(\hat{\tau}\) is a sum over a triangular array of independent, within-block-identically-distributed terms whose block sizes grow linearly in \(n\).
The finite second moment \(\sigma^2_{Y_0} < \infty\) placed on \(F\), together with the fixed-proportion counts, ensures Lindeberg's condition holds for this array (Vaart 1998, Proposition 2.27).
Hence, under the alternative with effect \(\tau\),
\[\frac{\hat{\tau} - \tau}{\text{SE}_\tau} \xrightarrow{d} N(0,1).\]
Along \(\tau_n = h/\sqrt{n}\), the permutation quantile converges in probability to \(z_{\alpha/2}\,\text{SE}_\tau\), since the \(\tau_n^2\)-inflation of the permutation variance is \(O(1/n)\) relative to \(\text{SE}^2\); contiguity then gives the stated limit.

So the standardized statistic has non-centrality parameter \(\delta = \tau/\text{SE}_\tau\) and the power formula follows. \(\square\)

\subsection{Power of Optimally Weighted RI (Oracle)}\label{proof:power-ri-optimal-oracle}

The optimally weighted estimator combines pair and singleton estimates:
\[\hat{\tau}_{\text{oracle}} = w^* \cdot \hat{\tau}_{\text{pair}} + (1-w^*) \cdot \hat{\tau}_{\text{single}}\]

\textbf{Step 1: Variances of component estimators.} From the proof of Proposition 1, the variance of the pair estimator is:
\[V_{\text{pair}} = \text{Var}(\hat{\tau}_{\text{pair}}) = \frac{2\sigma^2_{Y_0}(1-r)}{n(1-q)^2}\]

For singletons,
conditional on the attrition pattern, \(k_1 = k_0 = nq(1-q)\) treated and control singletons, permuted with fixed group sizes (as in the proof of Proposition 3), the variance is: \[V_{\text{single}} = \text{Var}(\hat{\tau}_{\text{single}}) = \frac{2\sigma^2_{Y_0}}{nq(1-q)}.\]

\textbf{Step 2: Inverse-variance weighting.} For two independent, unbiased estimators with variances \(V_1\) and \(V_2\), the minimum variance linear combination has weight:
\[w^* = \frac{1/V_1}{1/V_1 + 1/V_2}\]

and achieves variance:
\[\text{Var}(w^*\hat{\tau}_1 + (1-w^*)\hat{\tau}_2) = \frac{1}{1/V_1 + 1/V_2}\]

\textbf{Step 3: Apply to our setting.} Computing the optimal weight:
\[w^* = \frac{1/V_{\text{pair}}}{1/V_{\text{pair}} + 1/V_{\text{single}}} = \frac{n(1-q)^2/(2\sigma^2_{Y_0}(1-r))}{n(1-q)^2/(2\sigma^2_{Y_0}(1-r)) + nq(1-q)/(2\sigma^2_{Y_0})}\]

Simplifying (cancel \(\sigma^2_{Y_0}\) and factor out \(n(1-q)/2\)):
\[w^* = \frac{(1-q)/(1-r)}{(1-q)/(1-r) + q} = \frac{(1-q)}{(1-q) + q(1-r)}\]

\textbf{Step 4: Variance of optimal estimator.} The variance is:
\[\text{Var}(\hat{\tau}_{\text{oracle}}) = \frac{1}{1/V_{\text{pair}} + 1/V_{\text{single}}} = \frac{1}{\frac{n(1-q)^2}{2\sigma^2_{Y_0}(1-r)} + \frac{nq(1-q)}{2\sigma^2_{Y_0}}}\]

Factor out \(\frac{n(1-q)}{2\sigma^2_{Y_0}}\):
\[= \frac{2\sigma^2_{Y_0}}{n(1-q)\left[\frac{(1-q)}{(1-r)} + q\right]} = \frac{2\sigma^2_{Y_0}(1-r)}{n(1-q)[(1-q) + q(1-r)]}\]

\textbf{Step 5: Power calculation.} The oracle estimator \(\hat\tau_{\text{oracle}}\) is a fixed linear combination \(w^*\hat\tau_{\text{pair}} + (1-w^*)\hat\tau_{\text{single}}\) of the pair and singleton difference in means, each a sum of i.i.d. contributions drawn from \(F\) (as in the proof of Proposition 3). By the same CLT, \((\hat\tau_{\text{oracle}} - \tau)/\text{SE}_{\text{oracle}} \xrightarrow{d} N(0,1).\)
Along \(\tau_n = h/\sqrt{n}\), the permutation quantile converges in probability to \(z_{\alpha/2}\,\text{SE}_{\text{oracle}}\), since the \(\tau_n^2\)-inflation of the permutation variance is \(O(1/n)\) relative to \(\text{SE}^2\); contiguity then gives the stated limit.
So under the alternative with effect \(\tau\) the standardized statistic has non-centrality parameter \(\delta = \tau/\text{SE}_{\text{oracle}}\) and the power formula follows. \(\square\)

\subsection{Power of Optimally Weighted RI (Feasible)}\label{proof:power-ri-optimal-feasible}

The feasible estimator uses the same form as the oracle but with estimated variances:
\[\hat{\tau}_{\text{feasible}} = \hat{w}^* \cdot \hat{\tau}_{\text{pair}} + (1-\hat{w}^*) \cdot \hat{\tau}_{\text{single}}\]
where \(\hat{w}^* = \frac{(1-\hat{q})}{(1-\hat{q}) + \hat{q}(1-\hat{r})}\).

\textbf{Step 1: Decomposition.} Write
\[\hat{\tau}_{\text{feasible}} = \hat{\tau}_{\text{oracle}} + (\hat{w}^* - w^*)(\hat{\tau}_{\text{pair}} - \hat{\tau}_{\text{single}}).\]
Then
\[\text{Var}(\hat{\tau}_{\text{feasible}}) = \text{Var}(\hat{\tau}_{\text{oracle}}) + \text{Var}(R_n) + 2\,\text{Cov}(\hat{\tau}_{\text{oracle}}, R_n), \quad R_n := (\hat{w}^* - w^*)(\hat{\tau}_{\text{pair}} - \hat{\tau}_{\text{single}}).\]

\textbf{Step 2: Orders.} Standard asymptotics for sample variances and the delta method give \(\hat{w}^* - w^* = O_p(n^{-1/2})\), and \(\hat{\tau}_{\text{pair}} - \hat{\tau}_{\text{single}} = O_p(n^{-1/2})\) (both components are unbiased for \(\tau\), so the difference is mean-zero with variance \(V_{\text{pair}} + V_{\text{single}} = O(1/n)\)). Hence \(\text{Var}(R_n) = O(1/n^2)\). By Cauchy--Schwarz,
\[|\text{Cov}(\hat{\tau}_{\text{oracle}}, R_n)| \le \sqrt{\text{Var}(\hat{\tau}_{\text{oracle}})\,\text{Var}(R_n)} = O(n^{-1/2}) \cdot O(n^{-1}) = O(n^{-3/2}).\]

\textbf{Step 3: Rate.} Therefore \(\text{SE}_{\text{feasible}}^2 = \text{SE}_{\text{oracle}}^2 + O(n^{-3/2})\), the cross term being the binding one. This bound is not sharp; it is what Cauchy--Schwarz delivers without further moment assumptions.

\textbf{Step 4: Power approximation.}

For large \(n\), \(\hat\tau_{\text{feasible}}\) has the same limiting sampling distribution as \(\hat\tau_{\text{oracle}}\): by Step 3 the two share the same variance up to \(O(n^{-3/2})\), and \(\hat w^*\) is consistent for \(w^*\), so the CLT of Proposition 3 applies to \(\hat\tau_{\text{feasible}}\) as well. Under the alternative, \[\frac{\hat{\tau}_{\text{feasible}}}{\text{SE}_{\text{feasible}}} \approx \frac{\tau + O_p(1/\sqrt{n})}{\text{SE}_{\text{oracle}}\sqrt{1 + O(n^{-1/2})}} \approx \frac{\tau}{\text{SE}_{\text{oracle}}} + O_p(1/\sqrt{n})\]

The power formula follows by continuity of \(\Phi(\cdot)\). \(\square\)

\subsection{Proof of Proposition 6 (Exact Size Control)}\label{proof:exact-size-control}

\textbf{Standard RI result:} Under \(H_0: Y_{p,g}(1) = Y_{p,g}(0)\) for all \((p,g)\), observed outcomes \(Y_{p,g} = Y_{p,g}(0)\) are fixed and do not depend on treatment assignment \(D_{p,g}\). The permutation distribution therefore equals the randomization distribution, guaranteeing size \(\alpha\).

\textbf{Extension to attrition:} Let \(A_{p,g} \in \{0,1\}\) indicate whether unit \(g\) in pair \(p\) attrits (1) or is observed (0). We assume:
\[A_{p,g} \perp D_{p,g} \mid \mathbf{Y}_p\]

That is, attrition is independent of which treatment was assigned within a pair, conditional on the pair's potential outcomes.

\textbf{Key implication:} Under \(H_0\) where \(Y_{p,g}(1) = Y_{p,g}(0)\), the set of observed units \(\{(p,g): A_{p,g} = 0\}\) and their outcomes \(\{Y_{p,g}: A_{p,g} = 0\}\) are identical across all permutations of \(D\) within pairs. Each permutation \(D^{(\pi)}\) operates on the same set of observed values, merely reassigning which unit in each pair is labeled ``treated.''

\textbf{Formal argument:} Fix the attrition pattern \(\mathbf{A} = (A_{1,1}, A_{1,2}, \ldots, A_{n,1}, A_{n,2})\) and potential outcomes \(\mathbf{Y}(0) = (Y_{1,1}(0), Y_{1,2}(0), \ldots, Y_{n,1}(0), Y_{n,2}(0))\). Under \(H_0\) and conditional independence:
\[\mathbb{P}(|\hat{\tau}(D)| > c \mid \mathbf{Y}(0), \mathbf{A}) = \frac{1}{|\mathcal{D}|}\sum_{D^{(\pi)} \in \mathcal{D}} \mathbb{1}\{|\hat{\tau}(D^{(\pi)})| > c\}\]
where \(\mathcal{D}\) is the set of all treatment assignments consistent with pairwise randomization, and \(\hat{\tau}(D)\) is computed using only observed units \(\{(p,g): A_{p,g} = 0\}\).

Setting \(c = c_\alpha\) such that the right-hand side equals \(\alpha\) gives the rejection threshold. By construction, \(\mathbb{P}(\text{reject} \mid H_0) = \alpha\) exactly (up to permutation discreteness).

\textbf{Note on the attrition assumption:} The assumption \(A_{p,g} \perp D_{p,g} \mid \mathbf{Y}_p\) is weaker than requiring attrition to be completely random. It allows attrition to depend on a unit's own potential outcomes and on its partner's potential outcomes, but requires that conditional on the pair's full potential outcome vector, the realized treatment assignment does not affect attrition probability. This is not the missing-at-random assumption of the missing-data literature: attrition may depend on unobserved potential outcomes, which is missingness not at random in that taxonomy. What the assumption rules out is only dependence of attrition on the realized treatment assignment; we refer to it as attrition exogeneity with respect to assignment. It is the no-covariate analogue of the conditional independence assumption in Bai (2022) (Online Appendix C.3), which conditions on baseline covariates rather than on the pair's potential-outcome vector.

\textbf{Extension to feasible weights:} The exact size result extends to the feasible weighted procedure when nuisance parameters are recomputed for each permutation. Specifically, when weights \(\hat{w}^*(\pi)\) are computed separately for each permutation \(\pi\):

Under \(H_0: Y_{p,g}(1) = Y_{p,g}(0)\) for all \({p,g}\), the observed outcomes \(\{Y_{p,g}: A_{p,g} = 0\}\) are fixed and equal to \(\{Y_{p,g}(0): A_{p,g} = 0\}\). For each permutation \(\pi\):

\begin{itemize}
\tightlist
\item
  The set of units labeled as control, \(\mathcal{C}(\pi) = \{(p,g): D^{(\pi)}_{p,g} = 0, A_{p,g} = 0\}\), varies with the permutation
\item
  The control group variance \(\hat{\sigma}_{Y_0}^2(\pi) = \frac{1}{|\mathcal{C}(\pi)|-1}\sum_{(p,g) \in \mathcal{C}(\pi)} (Y_{p,g} - \bar{Y}_{\mathcal{C}(\pi)})^2\) depends on which units are in \(\mathcal{C}(\pi)\)
\item
  The within-pair variance \(\hat{\sigma}_{WP}^2(\pi) = \frac{1}{m-1}\sum_{p:\,\text{complete}}(\Delta_p^{(\pi)} - \hat{\tau}_{\text{pair}}(\pi))^2\), with \(\Delta_p^{(\pi)} = Y_{p,T(\pi)} - Y_{p,C(\pi)}\), is recomputed for each permutation; exactness does not require it to be permutation-invariant, only that it be a deterministic function of \(D^{(\pi)}\) and the fixed observed outcomes
\item
  The estimated correlation \(\hat{r}(\pi) = 1 - \hat{\sigma}_{WP}^2/(2\hat{\sigma}_{Y_0}^2(\pi))\) varies across permutations
\item
  The optimal weight \(\hat{w}^*(\pi) = (1-\hat{q})/[(1-\hat{q}) + \hat{q}(1-\hat{r}(\pi))]\) varies across permutations
\end{itemize}

The key is that under the sharp null, the joint distribution of \((\hat{\tau}(\pi), \hat{w}^*(\pi))\) over all permutations \(\pi \in \mathcal{D}\) equals their joint randomization distribution. Therefore, the permutation distribution of the feasible weighted test statistic \(\hat{\tau}_{\text{feasible}}(\pi) = \hat{w}^*(\pi) \hat{\tau}_{\text{pair}}(\pi) + (1-\hat{w}^*(\pi)) \hat{\tau}_{\text{single}}(\pi)\) equals its randomization distribution, ensuring:
\[\mathbb{P}(|\hat{\tau}_{\text{feasible}}(D)| > c_\alpha \mid H_0, \mathbf{Y}(0), \mathbf{A}) = \alpha\]
exactly, where \(c_\alpha\) is the \(\alpha\)-quantile of the permutation distribution.

In contrast, if weights were computed once from the observed data and held fixed across permutations, the permutation distribution would not account for sampling variability in \(\hat{\sigma}_{Y_0}^2\) induced by the random treatment assignment, breaking the equality between the permutation and randomization distributions. \(\square\)

\subsection{Proof of Proposition 7 (Dominance of optimally weighted RI)}\label{proof:dominance-optimal-ri}

We prove dominance by comparing non-centrality parameters using the power formulas in Propositions 1-4.

\textbf{Part (a): Oracle RI dominates paired \(t\)-test.}

From Propositions 1 and 4, oracle RI has greater power than paired FE when:
\[\frac{\tau}{\text{SE}_{\text{oracle}}} > \frac{\tau\sqrt{n(1-q)^2}}{\sqrt{2\sigma^2_{Y_0}(1-r)}}\]

Squaring both sides (assuming \(\tau > 0\)):
\[\frac{1}{\text{SE}_{\text{oracle}}^2} > \frac{n(1-q)^2}{2\sigma^2_{Y_0}(1-r)}\]

From Proposition 4:
\[\frac{1}{\text{SE}_{\text{oracle}}^2} = \frac{n(1-q)^2}{2\sigma^2_{Y_0}(1-r)} + \frac{nq(1-q)}{2\sigma^2_{Y_0}}\]

Therefore:
\[\frac{1}{\text{SE}_{\text{oracle}}^2} - \frac{n(1-q)^2}{2\sigma^2_{Y_0}(1-r)} = \frac{nq(1-q)}{2\sigma^2_{Y_0}} > 0\]
whenever \(q > 0\). Thus oracle RI strictly dominates the paired \(t\)-test for any positive attrition rate. \(\square\) (Part a)

\textbf{Part (b): Oracle RI dominates two-sample \(t\)-test.}

From Proposition 2, the two-sample \(t\)-test has power:
\[\text{Power}_{\text{Mean Diff}} = \Phi(\lambda(\delta - z_{\alpha/2})) + \Phi(\lambda(-\delta - z_{\alpha/2}))\]

From Proposition 4, the oracle optimally weighted RI has power:
\[\text{Power}_{\text{oracle}} = \Phi\left(\frac{\tau}{\text{SE}_{\text{oracle}}} - z_{\alpha/2}\right) + \Phi\left(-\frac{\tau}{\text{SE}_{\text{oracle}}} - z_{\alpha/2}\right)\]

\textbf{Step 1: Simplify the \(t\)-test non-centrality parameter.}

From Proposition 2, \(\delta = \tau\sqrt{n(1-q)}/\sqrt{2\sigma^2_{Y_0}}\) and \(\lambda = \sqrt{\text{SE}_{\text{assumed}}^2/\text{SE}_{\tau}^2}\) with \(\text{SE}_{\text{assumed}}^2 = 2\sigma^2_{Y_0}/[n(1-q)]\).

Therefore:
\[\lambda\delta = \frac{\tau\sqrt{n(1-q)}}{\sqrt{2\sigma^2_{Y_0}}} \cdot \sqrt{\frac{2\sigma^2_{Y_0}/[n(1-q)]}{\text{SE}_{\tau}^2}} = \frac{\tau}{\sqrt{\text{SE}_{\tau}^2}} = \frac{\tau}{\text{SE}_{\tau}}\]

where \(\text{SE}_{\tau}^2\) is from Proposition 3 (the variance of the simple difference in means under pairwise randomization).

\textbf{Step 2: Show \(\text{SE}_{\text{oracle}}^2 \leq \text{SE}_{\tau}^2\).}

From Proposition 3, the simple difference in means can be decomposed as:
\[\bar{Y}_T - \bar{Y}_C = (1-q)\hat{\tau}_{\text{pair}} + q\hat{\tau}_{\text{single}}\]

This is a linear combination of \(\hat{\tau}_{\text{pair}}\) and \(\hat{\tau}_{\text{single}}\) with fixed weights \((1-q, q)\). From Proposition 4, the oracle weighted estimator uses optimal weights \(w^*\) chosen to minimize variance. Therefore:
\[\text{SE}_{\text{oracle}}^2 \leq \text{Var}[(1-q)\hat{\tau}_{\text{pair}} + q\hat{\tau}_{\text{single}}] = \text{SE}_{\tau}^2\]

\textbf{Step 3: Show \(\lambda \geq 1\) in our setting.}

From the definitions, \(\lambda^2 = \text{SE}_{\text{assumed}}^2/\text{SE}_{\tau}^2\).
\[\text{SE}_{\tau}^2 = \frac{2\sigma^2_{Y_0}[1-r + qr]}{n(1-q)}\]

Since \(\text{SE}_{\text{assumed}}^2 = 2\sigma^2_{Y_0}/[n(1-q)]\):
\[\lambda^2 = \frac{2\sigma^2_{Y_0}/(n(1-q))}{2\sigma^2_{Y_0}[1-r + qr]/(n(1-q))} = \frac{1}{1-r + qr}\]

\[= \frac{1}{1 - r + qr} = \frac{1}{1 - r(1-q)}\]

For \(0 \leq q < 1\) and \(0 \leq r < 1\): we have \(r(1-q) \geq 0\), thus \(1 - r(1-q) \leq 1\), so \(\lambda^2 \geq 1\), thus \(\lambda \geq 1\).

\textbf{Step 4: Conclude.}

The power function \(g(\text{NC}, \text{CV}) = \Phi(\text{NC} - \text{CV}) + \Phi(-\text{NC} - \text{CV})\) is:
- Strictly increasing in the non-centrality parameter NC (for NC \textgreater{} 0)
- Strictly decreasing in the critical value CV (for CV \textgreater{} 0)

Comparing the two procedures:
- Oracle RI: NC = \(\tau/\text{SE}_{\text{oracle}}\), CV = \(z_{\alpha/2}\)
- Two-sample \(t\): NC = \(\tau/\text{SE}_{\tau}\), CV = \(\lambda z_{\alpha/2}\)

From Step 2: \(\tau/\text{SE}_{\text{oracle}} \geq \tau/\text{SE}_{\tau}\) (Oracle has larger NC)

From Step 3: \(z_{\alpha/2} \leq \lambda z_{\alpha/2}\) (Oracle has smaller CV)

Therefore:
\[\text{Power}_{\text{oracle}} = g\left(\frac{\tau}{\text{SE}_{\text{oracle}}}, z_{\alpha/2}\right) \geq g\left(\frac{\tau}{\text{SE}_{\tau}}, \lambda z_{\alpha/2}\right) = \text{Power}_{\text{Mean Diff}}\]

Thus oracle RI has power at least as great as the two-sample \(t\)-test. \(\square\)

\textbf{Remark:} Equality holds when both \(\text{SE}_{\text{oracle}} = \text{SE}_{\tau}\) (population-fraction weights equal optimal weights) and \(\lambda = 1\) (no matching benefit). For generic parameters with informative matching (\(r > 0\)), the inequality is strict.

\subsection{Proof of Lemma (Optimally Weighted Estimator via Weighted Regression)}\label{proof:weighted-regression}

The regression specification includes pair-specific fixed effects \(\alpha_p\) for each complete pair and a common fixed effect \(\alpha_0\) for all singletons.

Throughout this proof, \(i\) indexes all dataset rows in the generic OLS expressions. We then specialize: complete-pair rows are written via the pair index \(p\) (members \(g\in\{1,2\}\)), and singleton rows via a flat index \(s = 1, \ldots, k_1 + k_0\) with treatment label \(D_s \in \{T,C\}\) and outcome \(Y_s\). The weighted OLS estimator is:
\[\hat{\beta} = \frac{\sum_{i=1}^N \tilde{w}_i \tilde{D}_i \tilde{Y}_i}{\sum_{i=1}^N \tilde{w}_i \tilde{D}_i^2}\]
where tildes denote deviations from within-group means (within-pair for complete pairs, within-singleton-group for singletons).

Let \(m\) denote the observed number of complete pairs, \(k_1\) the observed number of treated singletons, and \(k_0\) the observed number of control singletons, with \(m = n(1-q)^2\) and \(k_1 = k_0 = nq(1-q)\).

\textbf{Step 1: Contribution from complete pairs.}

After demeaning within each pair \(p\), the treated unit has \(\tilde{D}_{p,T} = \frac12\) and \(\tilde{Y}_{p,T} = (Y_{p,T} - Y_{p,C})/2\), while the control unit has \(\tilde{D}_{p,C} = -1/2\) and \(\tilde{Y}_{p,C} = -(Y_{p,T} - Y_{p,C})/2\).

The contribution from pair \(p\) to the numerator is:
\[\tilde{w}_{\text{pair}} \left[\frac{1}{2} \cdot \frac{Y_{p,T} - Y_{p,C}}{2} + \left(-\frac{1}{2}\right) \cdot \left(-\frac{Y_{p,T} - Y_{p,C}}{2}\right)\right] = \tilde{w}_{\text{pair}} \cdot \frac{Y_{p,T} - Y_{p,C}}{2}\]

Summing over all \(m\) pairs:
\[\text{Numerator}_{\text{pairs}} = \frac{\tilde{w}_{\text{pair}}}{2} \sum_{p:\, \text{complete}}(Y_{p,T} - Y_{p,C}) = \frac{\tilde{w}_{\text{pair}} m}{2} \hat{\tau}_{\text{pair}}\]

The contribution to the denominator from all pairs is:
\[\text{Denominator}_{\text{pairs}} = \sum_{p:\, \text{complete}}\tilde{w}_{\text{pair}} \left[\left(\frac{1}{2}\right)^2 + \left(-\frac{1}{2}\right)^2\right] = \frac{\tilde{w}_{\text{pair}} m}{2}\]

\textbf{Step 2: Contribution from singletons.}

To ensure exact correspondence with the direct difference-in-means estimator \(\hat{\tau}_{\text{single}} = \bar{Y}_{T,\text{single}} - \bar{Y}_{C,\text{single}}\), we assign weights that preserve symmetric deviations \(\tilde{D}_i = \pm 1/2\) after demeaning, regardless of whether \(k_1 = k_0\). Specifically, we set:
\[\tilde{w}_{T,\text{single}} = \frac{1-w^*}{k_1}, \quad \tilde{w}_{C,\text{single}} = \frac{1-w^*}{k_0}\]

With these weights, the weighted mean treatment indicator among singletons is:
\[\bar{D}_{\text{weighted}} = \frac{k_1 \cdot \frac{1-w^*}{k_1} \cdot 1 + k_0 \cdot \frac{1-w^*}{k_0} \cdot 0}{k_1 \cdot \frac{1-w^*}{k_1} + k_0 \cdot \frac{1-w^*}{k_0}} = \frac{1-w^*}{2(1-w^*)} = \frac{1}{2}\]

After demeaning, treated singletons have \(\tilde{D}_s = 1 - 1/2 = 1/2\) and control singletons have \(\tilde{D}_s = 0 - 1/2 = -1/2\).

The deviations \(\tilde{Y}_s = Y_s - \bar{Y}_{\text{weighted,single}}\) satisfy:
\[\sum_{s:D_s=T} \tilde{w}_{T,\text{single}} \tilde{Y}_s = \frac{1-w^*}{k_1} \sum_{s:D_s=T} (Y_s - \bar{Y}_{\text{weighted,single}}) = \frac{1-w^*}{2}(\bar{Y}_{T,\text{single}} - \bar{Y}_{C,\text{single}})\]

Similarly for control singletons. The contribution to the numerator is:
\[\text{Numerator}_{\text{single}} = \frac{1}{2} \cdot \frac{1-w^*}{2}\hat{\tau}_{\text{single}} + \left(-\frac{1}{2}\right) \cdot \left(-\frac{1-w^*}{2}\right)\hat{\tau}_{\text{single}} = \frac{1-w^*}{2}\hat{\tau}_{\text{single}}\]

The contribution to the denominator is:
\[\text{Denominator}_{\text{single}} = k_1 \cdot \frac{1-w^*}{k_1} \cdot \frac{1}{4} + k_0 \cdot \frac{1-w^*}{k_0} \cdot \frac{1}{4} = \frac{1-w^*}{2}\]

\textbf{Step 3: Combine to show equivalence to optimally weighted estimator.}

The weighted OLS estimator is:
\[\hat{\beta} = \frac{\frac{\tilde{w}_{\text{pair}} m}{2}\hat{\tau}_{\text{pair}} + \frac{1-w^*}{2}\hat{\tau}_{\text{single}}}{\frac{\tilde{w}_{\text{pair}} m}{2} + \frac{1-w^*}{2}}\]

We need to show this equals \(w^* \hat{\tau}_{\text{pair}} + (1-w^*) \hat{\tau}_{\text{single}}\).

\textbf{Step 4: Setting observation weights based on precision.}

We set pair observation weights proportional to precision:
\[\tilde{w}_{\text{pair}} = \frac{c_0}{1-r}\]
where \(c_0\) is a normalizing constant. The singleton weights from Step 2 already embody
the principle of equal weighting (precision \(\propto 1\)) through the construction
\(\tilde{w}_{T,\text{single}} = (1-w^*)/k_1\) and \(\tilde{w}_{C,\text{single}} = (1-w^*)/k_0\).

\textbf{Step 5: Verifying this produces the optimal estimator.}

From Step 3, the weighted regression gives:
\[\hat{\beta} = \frac{\frac{c_0m}{2(1-r)}\hat{\tau}_{\text{pair}} + \frac{1-w^*}{2}\hat{\tau}_{\text{single}}}{\frac{c_0m}{2(1-r)} + \frac{1-w^*}{2}}\]

For this to equal \(w^* \hat{\tau}_{\text{pair}} + (1-w^*) \hat{\tau}_{\text{single}}\), we need the coefficient on \(\hat{\tau}_{\text{pair}}\) to equal \(w^*\):
\[\frac{c_0m/(1-r)}{c_0m/(1-r) + (1-w^*)} = w^*\]

Solving for \(c_0\):
\[c_0m/(1-r) = w^*[c_0m/(1-r) + (1-w^*)]\]
\[c_0m/(1-r)(1-w^*) = w^*(1-w^*)\]
\[c_0 = \frac{w^*(1-r)}{m}\]

Therefore \(\tilde{w}_{\text{pair}} = w^*/m\), confirming that:
\[\hat{\beta} = w^* \hat{\tau}_{\text{pair}} + (1-w^*) \hat{\tau}_{\text{single}} = \hat{\tau}_{\text{oracle}} \quad \square\]

\textbf{Remark on implementation.} The observation weights can be implemented in two equivalent ways: (1) use \(\tilde{w}_{\text{pair}} = w^*/m\) and \(\tilde{w}_{T,\text{single}} = (1-w^*)/k_1\), \(\tilde{w}_{C,\text{single}} = (1-w^*)/k_0\) as derived above, or (2) use any weights satisfying \(\tilde{w}_{\text{pair}} \propto 1/(1-r)\) and \(\tilde{w}_{\text{single}} \propto 1\), since the regression automatically normalizes to produce the same coefficient. The proportionality to precision is what matters, not the absolute scale.

\end{document}